\documentclass{article}

\usepackage{PRIMEarxiv}

\usepackage[utf8]{inputenc} 
\usepackage[T1]{fontenc}    
\usepackage{hyperref}       
\usepackage{url}            
\usepackage{booktabs}       
\usepackage{amsfonts}       
\usepackage{nicefrac}       
\usepackage{microtype}      
\usepackage{lipsum}
\usepackage{fancyhdr}       
\usepackage{graphicx}       
\graphicspath{{media/}}     
\usepackage{changepage}
\usepackage{tabularx}
\usepackage{makecell}
\usepackage{pgfplots}
\pgfplotsset{width=20cm,compat=1.8}
\usepackage{subcaption}
\usepackage{pgf-pie}

\newcommand*\justifyy{%
	\fontdimen2\font=0.4em
	\fontdimen3\font=0.2em
	\fontdimen4\font=0.1em
	\fontdimen7\font=0.1em
	\hyphenchar\font=`\-
}
\newcommand{\app}[1]{{\tt \justifyy{\textbf{#1}}}}
\title{Evaluating Plant Disease Detection Mobile Applications: Quality and Limitations
}

\author{
  Ayesha Siddiqua \\
  Department of Computer Science and Engineering\\
  Central Women's University\\
  Dhaka, 1203, Bangladesh\\
  \textit{ayesha.siddiqua@cwu.edu.bd}\\
   \And
  Muhammad Ashad Kabir\\
  School of Computing, Mathematics and Engineering\\
  Charles sturt Univerity\\
  Bathurst, NSW 2795, Australia\\
  \textit{akabir@csu.edu.au}
   \And
   Tanzina Ferdous\\
   Department of Computer Science and Engineering\\
   City University\\
   Dhaka, Bangladesh\\
   \textit{tanzina@cityuniversity.edu.bd}\\
   \And
   Israt Bintea Ali\\
   Divine IT Limited\\
   Dhaka, Bangladesh\\
   \textit{israt@divine-it.net}\\
   \And
   Leslie A. Weston\\
   Gulbali Institute for Agriculture, Water and Environment\\
   Charles Sturt University\\
   Wagga Wagga, NSW, 2678, Australia\\
   \textit{leweston@csu.edu.au}
}

\begin{document}
\maketitle

\begin{abstract}
In this technologically advanced era, with the proliferation of artificial intelligence, many mobile apps are available for plant disease detection, diagnosis and treatment, each with a variety of features. These apps need to be categorized and reviewed following a proper framework that ensures their quality. This study aims to present an approach to evaluating plant disease detection mobile apps, this includes providing ratings of distinct features of the apps and insights into the exploitation of artificial intelligence used in plant disease detection. The applicability of these apps for pathogen or disease detection, identification, and treatment will be assessed along with significant insights garnered. For this purpose, plant disease detection apps were searched in three prominent app stores (the Google Play store, Apple App store, and Microsoft store) using a set of keywords. A total of 606 apps were found and from them 17 relevant apps were identified based on inclusion and exclusion criteria. The selected apps were reviewed by three raters using our devised app rating scale. To validate the raters' agreement on the ratings, inter-rater reliability is computed alongside their intra-rater reliability, ensuring their rating consistency. Also, the internal consistency of our rating scale was evaluated against all selected apps. User comments from the app stores are collected and analyzed to understand their expectations and views. Following the rating procedure, most apps earned acceptable ratings in software quality characteristics such as aesthetics, usability and performance, but gained poor ratings in AI-based advanced functionality, which is the key aspect of this study. However, most of the apps cannot be used as a complete solution to plant disease detection, diagnosis and treatment. Only one app, \app{Plantix - your crop doctor}, could successfully identify plants from images, detect diseases, maintain a rich plant database, and suggest potential treatment for the disease presented. It also provides a community where plant lovers can communicate with each other to gain additional benefits. In general, all existing apps need to improve functionalities, user experience, and software quality. Therefore, a set of design considerations has been proposed for future app improvements.
\end{abstract}


\section{Introduction}

Plant diseases are a naturally occurring phenomenon limiting plant growth, development and reproduction~\cite{agrios2005plant}. Plant diseases and the pathogens causing them are a direct threat to the global economy and food security~\cite{leafDoctor}. Plant diseases are typically caused by pests, including viral, bacterial, and fungal-like organisms that hinder the normal growth of plants and cause variations in their vital functions~\cite{plantdisease}. For example, Leaf Blast, Brown Spot, Leaf Blight, Sheath Rot, and Stem Rot are common rice diseases which reduce grain quality and destroy the majority of rice seedlings when badly infested~\cite{phadikar2008rice}. \textit{Phytophthora infestans} or Late Blight, and \textit{Alternaria solani} or Early Blight, are two well-characterised diseases that cause significant yield losses in potatoes~\cite{islam2017detection}. Stripe Rust, Powdery Mildew, Leaf Rust, and other diseases harm winter wheat in central Asia, causing significant grain yield losses~\cite{huang2014new,article}. 

In contrast, severe grape leaf diseases like Black Rot and Powdery Mildew have a significant impact on grape yield and quality during the growth process, particularly in wet or moist humid conditions~\cite{meunkaewjinda2008grape,sannakki2013diagnosis, 6707647, Xie}. Soybean exhibits numerous leaf diseases such as Anthracnose, Bacterial Blight, Soybean Mosaic Virus, Copper Phytotoxicity, Charcoal Rot, Leaf Cercospora, Rhizoctonia Aerial Blight, and Downy Mildew. Plant pathogens causing disease have become a great menace when it comes to crops or other cultivated plat materials as our livelihood is dependent on them~\cite{KARLEKAR2020105342}. Therefore, identifying plant diseases is crucial for avoiding long-term or periodic harm to harvested produce, and living landscape plants~\cite{al2011fast}. Early diagnosis of plant diseases allows for the implementation of preventive measures and the reduction of economic and production damage~\cite{islam2017detection}. For sustainable agriculture, plant health observation and predictive maintenance are essential for reducing losses related to agricultural commodity yields and quality~\cite{khirade2015plant}.

A traditional way of treating plant diseases is through the direct consultation of farmers and horticulturists with plant pathologists or outreach specialists in person. Experts physically examine different parts of the symptomatic plant or test the soil to evaluate soil properties, nutrients or pH and treatments are then suggested. However, the traditional method of diagnosing plant diseases can be unreliable and limited by available expertise~\cite{cassava,Dutot}. Continuous plant surveillance by land managers is not always practical or cost-effective~\cite{francis2016identification}. For this reason, mobile and other technologies can provide a low-cost, high-precision alternative~\cite{petrellis2017mobile}. There are various symptoms that can be used to identify plant pathogens and related diseases~\cite{riley2002plant}. Modern technology has been brought into the picture to enhance the diagnostic process. This sector is modernized now with the advent of technology as smartphones have made their way into everyone's lives, including those of rural farmers. As a result, many applications related to plants and their pathogens/diseases have been built on many platforms in recent years~\cite{BEDI202190,agriengineering3030032,ict}.

Existing plant and plant disease-related apps have a broad range of functionalities, such as identifying of plants and diseases, disease severity estimation, biological knowledge, agricultural solutions, marketplace parameters, and more. Plant identification and disease detection are the most critical and beneficial features of the apps as stakeholders take the required steps based on the result of plant disease detection. To provide these functionalities, image processing techniques, the grading method, artificial intelligence (AI), machine learning (ML), and deep learning (DL) techniques have been applied, supported by a large image data-set of plant varieties and their diseases~\cite{singh2018deep, 9399342,cottonfarming, petrellis2019plant, ashok2020tomato, weizheng2008grading, 5697452,arsenovic2019solving,agriculture12010009,PICON2019280,YUAN202248}.

Several apps claim to effectively recognize and treat plant diseases on various platforms, but unfortunately, there is no accreditation of apps, which is why individuals may face difficulty selecting the best app. Also, verification of the functionality does not mean the app is user-friendly and effective, as there are many evaluation metrics, such as user interface, error rates, response time, transparency, etc., other than user ratings. So, plant disease detection apps require a thorough analysis using a proper methodology. 

To the best of our knowledge, no such structured study has yet been conducted on plant disease detection, severity estimation and treatment related mobile apps. A brief description of agricultural apps and plant disease apps has been provided by researchers~\cite{che2022mobile,sibanda2021systematic,Agrisurvey, multimediaCrop}, however, they did not focus on plant disease detection, diagnosis and care related criteria. In contrast, this study aims to review plant disease detection applications from three different platforms (the Google Play store, Apple App store and Microsoft store) by giving equal importance to the disease identification specific functionalities and software quality ensuring criteria. In this paper, we have made the following five major contributions: 

\begin{enumerate}
    \item We have examined the existing apps accessible in the three major app stores (i.e., Google Play store, Apple App store, and Microsoft store).
    \item We have devised an app rating scale for analyzing plant disease detection apps by adopting and extending existing rating scales.
    \item We have evaluated the selected apps through our devised app rating scale and and identified their design issues.
    \item We have analyzed app store user comments to better understand users' expectations and perspectives.
    
    \item We have provided design guidelines that emphasize using artificial intelligence for better plant disease detection app development.
\end{enumerate}

The paper is organized as follows.
Section~\ref{sec:methodology} describes the methodology and Section~\ref{sec:results} presents our evaluation results. Section~\ref{sec:discussions} discusses the limitations of the reviewed apps and design considerations for better app development. Finally, Section~\ref{sec:conclusion} concludes the paper. 

\section{Methodology}\label{sec:methodology}

\subsection{App search strategy}
\begin{figure}[htbp]
    \begin{center}
     \includegraphics[width=1\textwidth]{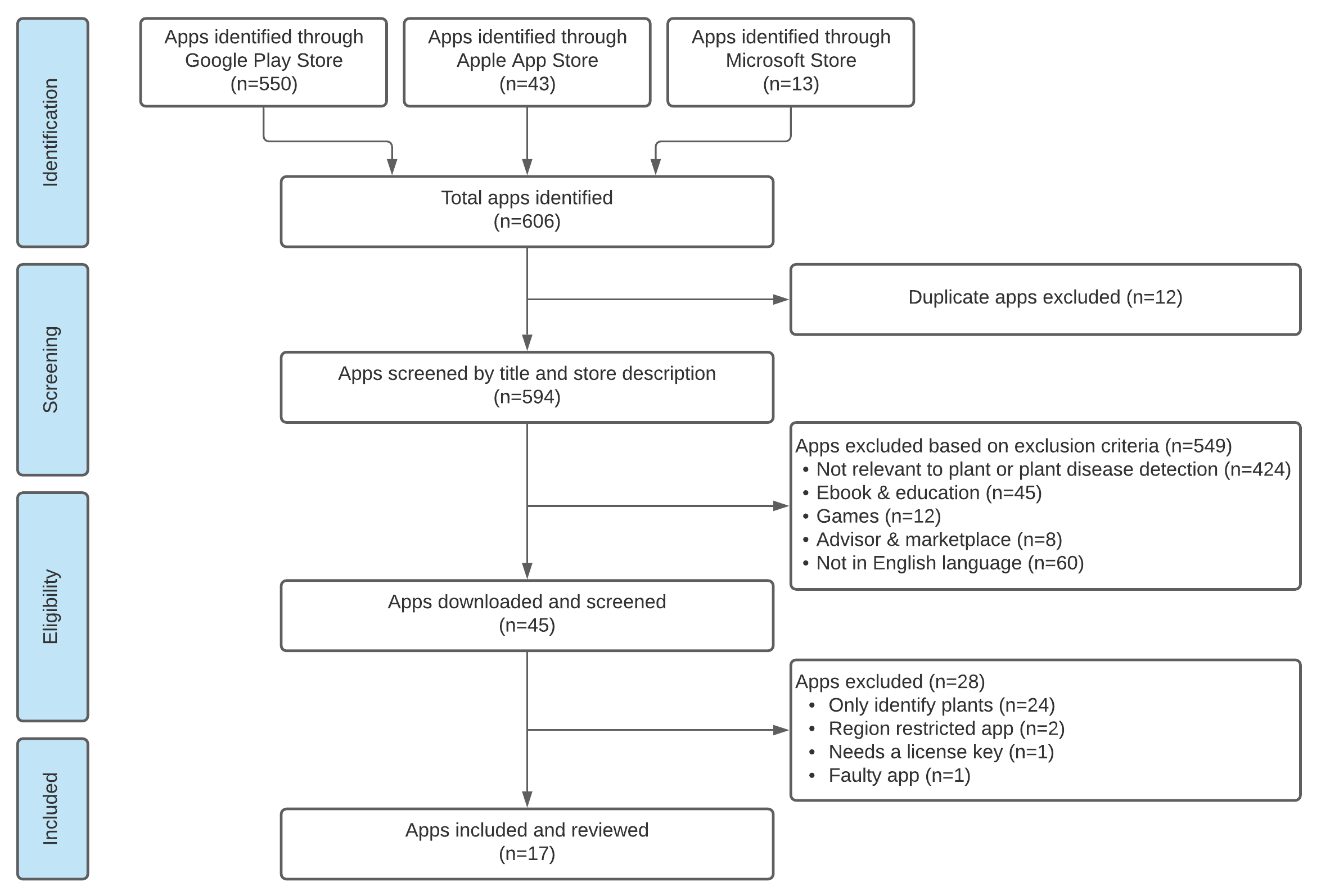}
     \caption{PRISMA diagram of study}
     \label{prisma16}       
    \end{center}
\end{figure}

Apps were searched in the three most popular app stores: Google Play, Apple App and Microsoft. A keyword-based search process has been employed following a similar approach used in previous studies~\cite{rivera2016mobile,Kabir2021}. The keywords that were used included: ``plant disease", ``leaf disease", ``plant disease detection", and ``leaf disease detection".
To maintain transparency and clarity in reporting, as well as the opportunity for other researchers to recreate the systematic search procedure for apps, a structured strategy PRISMA was followed~\cite{tricco2018prisma,liberati2009prisma}. The complete process of app searching, screening, and selection is presented using the PRISMA diagram in Fig.~\ref{prisma16}. After removing the duplicate apps, multiple screenings were performed to remove the redundancies and irrelevancies by mutual consensus of the researchers of this paper. The selected apps were further assessed quality-wise to avoid any bias. This extensive search procedure was executed twice to ensure no relevant apps are missed in the screening process.

Inclusion and exclusion criteria were defined to select relevant apps. In the screening process, an app was included if it is related to plant disease detection and mentions this feature in the \textit{description} or \textit{about} section of the app in the respective app store (inclusion criteria). The exclusion criteria were: (i) apps that are not relevant to plant or plant disease detection; (ii) apps that are only information based or educational about plant diseases; (iii) apps under the category of games since they do not have any relevance with the scope of our review; (iv) apps that provide a marketplace for farming essentials such as fertilizers or pesticides, or provide expert farming consultancy or advice; and (v) apps that are not in English as other languages are not comprehensible to us.

The investigators collaborated on the inquiry, screening, and retrieval phases of the app searching and collection processes. Each kept a separate list of apps they discovered in app stores using the finalized procedure's inclusion and elimination parameters. The investigators used their personal devices to decide which apps are suitable for selection. Some problems were faced during the accumulation of the individual app lists, for instance an app was excluded by one investigator but added by another investigator. Disagreements among the investigators were settled by discussion before consensus was achieved. The resulting app lists were combined to create the final list of apps for review (n=606). Multiple screenings were performed to reach a consensus among researchers to rule out apps that were irrelevant to the review. After removing duplicate apps (n=12), title and description-based screening was performed, which identified eligible apps (n=45) for the review, considering the exclusion criteria. Those 45 apps were installed and screened, and in that process, 28 apps were excluded for various reasons. This screening process resulted in selecting 17 apps for our review. 

\subsection{Raters} 
The selected apps were evaluated by three raters using our devised mobile app rating scale (see Section~\ref{sec:rating-scale}). The raters reviewed all the apps independently in a Google form and then information was collected in a Google sheet and processed later for assessing different parameters. The three expert raters include one software developer with over four years of mobile app development experience in a reputed company and two computer science graduates with two years of mobile apps development experience. In addition, the raters were under the guidance of a PhD graduate in software engineering with ten years of mobile app research experience. He also acted as a fourth rater whenever there was any discrepancy among the three raters and required to resolve the matter.

\subsection{App rating scale for plant disease detection}\label{sec:rating-scale}
A standardized approach was necessary to assess the mobile apps based on various aspects of the mobile application. Moreover, such an approach allowed us to assess and compare apps objectively, considering their functionality and different software quality criteria. The Mobile App Rating Scale (MARS) for mobile health apps~\cite{stoyanov2015mobile} is a popular rating scale that provides a multidimensional measure of the app quality assessment. We adopted and extended the existing app rating scales~\cite{Kabir2021,pritha2022smartphone,sabiha2022,stoyanov2015mobile,umars,finmars} for devising our app rating scale for plant disease detection. 

The set of sub-scales that were adopted and extended to our plant disease detection apps' rating scale from the mentioned tools was as follows: (i) app classification; (ii) aesthetics; (iii) general features; (iv) performance and efficiency; (v) usability; (vi) functionality (plant disease detection related); (vii) transparency; (viii) subjective quality; and (ix) the app’s perceived impacts on users. Some of the sub-scales for rating were adjusted (e.g., app classification) or newly defined (e.g., functionality) as needed to satisfy the objectives of plant disease detection apps. 
The rating scale was constructed using the app quality standards clustered around the domains, excluding the metadata portion. Each question was scored in a range of 1 to 5 using a Likert scale~\cite{wu2007empirical} or binary answer, depending on the type of question presented to determine the criterion. 

\subsubsection{App metadata}
Metadata provides basic information about the subject of interest.
Table \ref{Apps meta data} lists the metadata such as platform, country of origin, business model (free/paid), and the number of downloads for each of our evaluated apps. However, this information has not been used in the app scoring process. For our selected plant disease detection apps, metadata are extracted from the respective app stores. The iOS app store does not provide download numbers and thus 5 apps' download numbers are missing in the table. We have found 5 apps with 10K+ downloads. All the 5 iOS apps were pay to use, and the remaining 12 Android apps were free to download and use. 

\begin{table}[htbp]
    \centering
    \caption{Apps meta data.}
    \label{Apps meta data}
    \begin{tabularx}{\textwidth}{lcccc}
   \toprule
       App name  & Country of origin & Free/paid & Downloads & Platform \\
 \midrule
      Pestoz Idenify Plant Diseases & India & Free & 10K+ & Android\\
      
     AgroAI - Plant Diseases Diagnosis (Early Access) & Africa & Free & 10+ & Android\\
      
      Cropalyser & Netherlands & Free & 10K+ & Android\\
     
      PDDApp: Plant Disease Detection & Russia & Free & 1K+ & Android\\
      
      Leafy & India & Free & 100+ & Android\\
      
      PlantDoctor & India & Free & 1K+ & Android\\
      
      Plantix - your crop doctor & Germany & Free & 100K+ & Android\\
      
      Plant Disease Detector & Unknown & Free & 10+ & Android\\
      
      Riceye & Unknown & Free & 5+ & Android\\
      
      Cassava Plant Disease Identify & Unknown & Paid & -- & iOS\\
      
      Plants Disease Identification & Unknown & Paid & -- & iOS\\
      
      Garden Plants Diseases Detector & Unknown & Paid & -- & iOS\\
      
      Plant Disease Identifier & Unknown & Paid & -- & iOS\\
      
      Plant Diseases and Pests & Unknown & Paid & -- & iOS\\
      
      PlantifyDr & United States & Free & 10+ & Android and iOS\\
      
      Leaf Doctor & United States & Free & 10K+ & Android and iOS\\
      
      Agrio & United States & Free & 100K+ & Android and iOS\\
   \bottomrule
      \multicolumn{5}{l}{-- Apple app store does not provide download information}\\
    \end{tabularx}
\end{table}

\subsubsection{Aesthetics}\label{Aesthetics}
The aesthetics of an app plays an important role in a good user experience. The attractiveness or look of an interactive user interface is referred to as visual aesthetics, which is important for the functionality and usability of an app since it affects users' first impressions. Users' emotions and emotional reactions can be generated by the visual components of technology products~\cite{silvennoinen2014experiencing}. It is a critical component in the user interface's effective usability, satisfaction, trust and credibility, and preference~\cite{von2018we,isO,zen2016assessing,tractinsky2000beautiful}. 

Existing plant disease detection apps have been developed for gardeners, students and producers. So, the buttons, icons, or menus on the screen must be appropriate and zoomable. The resolution of graphics must also be good. Producers from rural or urban areas use such apps, so the apps' visual information and design must be appropriate for the target audience. Thus, aesthetics was rated on these criteria for each app -- layout, graphics, visual appeal, and appropriateness for the target user group. 

\subsubsection{General features}\label{General}

Several general features are expected in a plant disease detection app, including data export option, login or sign-up, notification, multi-languages support, and on-board tutorial for a smooth user experience. Some of these features are more objective than subjective. An app must provide basic functionality, like social sharing and data exporting. These capabilities allow users to save information for later use and share it with others. Thus, the absence of these characteristics was rated as a low point. A premium subscription option is also an important aspect of an app as consumer spending on premium apps is expected to reach \$270 billion by 2025, over 2.5 times what it was in 2020~\cite{sensortower}.

A plant disease detection app must also provide an on-board tutorial so that a new gardener/user can understand clearly how to use the app (e.g., taking picture, etc.). If the app provides customization features, it helps users to set their preferred language and use it with more comfort. Notifications are designed to make it easier for users to remember new features or messages and improve the app's usage frequency. It can also alert users to take routine care of their plants or crops. Thus, all these above-mentioned criteria were included in the general features sub-scale of our rating scale.

\subsubsection{Performance and efficiency}\label{perform}
A plant disease detection app requires a compatible device to function properly. For example, the device must have a resonable camera to take high quality photos of plants or leaves and roots. The app should also offer a quick cure or suggestions for users facing a problem with their plants. This sub-scale includes CPU performance, memory usage, battery consumption, the extent of heating up of the device, smooth working UI components, whether the app crashes while running, etc. Thus, these criteria were included in our rating scale's the performance and efficiency sub-scale.

\subsubsection{Usability}\label{Usab}
The term ``usability" refers to the entire relationship between the user and the product~\cite{wei2015usability}. It directly impacts how users feel about an app and may help them become long-term users. Even if people run into issues, a solution should be simple to discover. The usability attribute analyzes how simple a system interface is to use~\cite{payne2015behavioral}. Usability testing is critical in determining if an app is of sufficient quality to attract the attention of its desired user groups~\cite{kallio2005usability}. We consider ease of use, ease of navigation between components, and whether interactions between components are intuitive in the usability sub-scale of our app rating scale.

\subsubsection{Functionality}\label{Func}
Functionality validates the goal of an app and its purpose. People search for apps in app stores using essential functionality as a keyword, demonstrating how significant this criterion is for an app~\cite{7961668}. 
The functionality assessment criteria for plant disease detection apps included plant identification, plant coverage, disease detection, infected area visualization, disease severity estimation, treatment and expert/community support.

To define the rating scale of those assessment criteria, we emphasize technological advancement, such as the adoption of artificial intelligence that provides automation.
Manual techniques in conventional planting processes cannot cover huge plantations or give critical early instructions in decision-making processes~\cite{miller2009plant}. As a result, it is important to create automated solutions that are practical, reliable, and cost-effective in monitoring plant health and delivering useful information to decision-makers, such as disease detection, estimating its severity and recommending the correct dose of pesticides in the treatment~\cite{mahlein2016plant}.
The functionality sub-scale criteria and the definitions of the rating scores to evaluate those criteria are presented in Table~\ref{func}. We discuss each of those assessment criteria below.

\begin{table}[htbp]
    \centering
    \caption{Functionality assessment criteria definition.}
    \label{func}
    \begin{tabularx}{\textwidth}{llll}
    \toprule
    Functionality assessment criteria & Rating 5  & Rating 3  & Rating 1\\
 \midrule
     Plant identification & Automatically from image & Manually &  Not at all \\
     Plant coverage & 10+ & 1 to 10 & 0\\
     Disease detection  & Automatically from image  & From questionnaire  & Not at all \\
     Visualize infected area  & Automatically  & Required user input  & Not at all\\
     Disease severity estimation  & Automatically  & Required user input  & Not at all\\
     Treatment  & Up to date suggestions  & Fixed suggestions  & Not at all\\
     Expert/community support  & Expert and community &  Expert or community & Not at all \\
   \bottomrule
    \end{tabularx}
\end{table}

There is a strong demand for automated plant identification that aids amateur users who lack specialized knowledge in plant science in identifying plant species~\cite{waldchen2018automated}. In recent years, we have witnessed ample research on artificial intelligent-based automatic plant identification from plant images~\cite{9399342,wang2017review}. The leaf is an appropriate organ for this identification system as it has several distinct features, such as shape, vein, and texture~\cite{wang2017review, nijalingappa2015plant}. Thus, leaf identification is critical in plant classification~\cite{agarwal2018plant, nijalingappa2015plant}. \cite{wang2017review} provided a thorough overview of image processing techniques for this automatic plant identification from the leaf along with some frequently used machine learning algorithms such as SVM, KNN, RF, K-Means, and Hypersphere classifiers. For large-scale plant classification, the deep learning method that uses a convolutional neural network (CNN) to generate discriminating features for plant classification has become more prominent recently~\cite{9399342,vanrecognition, bodhwani2019deep,FERENTINOS2018311}. Hence, a rating of 5 has been given for the automation of plant identification, which means the app will identify the plant from the given picture. A rating of 3 has been assigned for the manual selection of the plant as plant identification is crucial for disease detection. Here, users will select a plant from the given list of plants in the app, or they will have to mention the name of the plant. The lowest rating of 1 has been given to the lack of plant identification.

The range of different plants for pathogen diagnosis is also a critical aspect, as with the inclusion of various plants, apps can cover a much broader range of diseases. The range of plants was established by examining the apps and appropriate values were assigned on the rating scale. Thelarger the plant coverage, the higher its value on the rating scale and lowest value has been given for zero plant coverage which means app does not cover any single plant.

Deep learning is a popular machine learning technique for detecting plant disease due to its automatic prediction from large amounts of data with high accuracy. One such commonly used dataset is ``PlantVillage"~\cite{Plantvillage}, which contains 54K+ healthy and unhealthy leaf photos in 38 groups based on species and disease~\cite{geetharamani2019identification}. 
Using leaf images of healthy and sick plants, customized deep learning models based on unique convolutional neural network architectures have been constructed to diagnose of plant diseases~\cite{FERENTINOS2018311}. Many conventional deep learning architectures are paired with optimization and customization approaches to provide considerable accuracy using technology in plant disease detection~\cite{ABADE2021106125,BISCHOFF2021105922, FERENTINOS2018311}. The performance of several deep learning techniques has been studied and compared~\cite{LEE2020105220}. Therefore, it is expected that mobile apps should implement them to automate disease detection. Thus, the highest value of 5 has been assigned in the rating scale for automatic disease identification from the plant image. A rating of 3 has been assigned if any app identifies the disease through a questionnaire, and a rating of 1 has been given for the inability to detect disease.

After recognizing the plant disease, identifying the infected portion and its severity estimation are crucial tasks as they will inform the user about the disease’s threat level. Severity estimation refers to the precise measurement of the leaf area showing disease symptoms~\cite{WSPANIALY2020105701}. The unhealthy region of the plant leaves can be detected by the color space conversion, then masking the green pixels (as they indicate the healthy areas of leaves) and finally applying a threshold value to remove the green pixels so that the image contains only diseased areas~\cite{arivazhagan2013detection}.  A novel network, PD2SE-Net~\cite{LIANG2019518}, has been introduced to diagnose and estimate the severity of plant diseases. A U-Net architecture has also been utilized to segment sick leaf areas~\cite{WSPANIALY2020105701}. So, if the result can be gained automatically from the image for both the visualization and severity estimation of diseased areas, the highest value of 5 has been assigned on the rating scale. On the other hand, if the user needs to point out the diseased portion of a leaf before the result is obtained, it has been rated with a moderate value of 3. A value of 1 is assigned for the absence of these features.

Recognizing a critical issue is important, but solving the problem is even more important. The same may be said for apps for identifying plant diseases. After the disease has been identified, the user must know what to do for the cure. Apps can recommend a cure from their own solutions or link users to Google's solutions. Because the information provided by Google is recent, the highest rating of 5 has been assigned for it. A rating of 3 has been assigned for the fixed treatment information. 

Even after all the solutions, getting aid from professionals or sharing any sort of problem with fellow users feels fantastic. As a result, community support is a valuable resource, and users become more engaged to the app because of their sharing. It also serves as a link between users and specialists. So, the highest rating of 5 has been assigned for any sort of expert and community assistance, and a lower rating of 3 for solely community assistance.

\subsubsection{Transparency}\label{transp}

In any development, transparency refers to the open and honest sharing of a project's status in real-time. The transparency of mobile apps has been a significant concern for consumers in recent times~\cite{betzing2020impact}. Nowadays, software developers are dedicated to gaining and maintaining user trust, which is critical to the industry's growth and development. However, users are uninformed of how information is stored and disseminated because data transmission occurs in the background of apps~\cite{HotMobile}. Also, trust is earned by providing explicit descriptions of what each app performs and acting in accordance with those explanations. Consumers' sharing of data between apps will be available to all users, regardless of the app's ability to work or not work. Consumers will choose when and how much they can share with other people.
The following factors were examined while evaluating the transparency requirements of the selected apps: (1) validity of the information provided in the app store description; (2) authenticity of the publisher or developer and the app source; (3) evidence-based app trial; (4) determining whether the app is feasible to achieve the developer's stated goals; and (5) a general alert to ask users' permission before collecting personal or private data and location information~\cite{betzing2020impact}.

\subsubsection{Subjective quality}\label{subjq}
An app’s subjective quality is related to the user's opinion, satisfaction, dissatisfaction, or overall view of the app. 
To assess the subjective quality, the raters looked at how satisfied they are with an app, how likely they are to pay for it, and how they would rate it overall.
This sub-scale considers the user's willingness to pay for the app, the user's recommendation for the app, and the user's rating of the app. 

\subsubsection{Perceived impact of users}\label{perimp}
The use of an app can influence a person’s attitude toward relevant areas of their life. Considering plant disease detection apps can influence a user's knowledge of plant diseases, a well-informed app can make a person’s gardening life easier and improve a user's knowledge about relevant areas. Furthermore, such apps can influence people to garden or gain a better understanding of plant care issues.
Therefore, the criteria to evaluate an app's perceived impact on its users are defined as: (1) improve awareness of the necessity of addressing plant diseases; (2) increase knowledge or understanding of plant diseases; (3) change people's minds about how to improve plantation and care; (4) improve intentions or motivation to address gardening; (5) stimulate additional plant management help seeking; and (6) increase plant disease treatment.

\section{Results}\label{sec:results}

\subsection{Internal consistency of our app rating scale}
Internal consistency describes the extent to which all the items in a test measure the same concept or construct (in our case the questions/items used in a sub-scale/assessment criteria), such that the items are consistent with one another and measuring the same thing~\cite{christmann2006robust}. We used Cronbach's alpha, the most popular means of calculating internal consistency~\cite{cronbach1951coefficient}. Cronbach's alpha ($\alpha$) reliability coefficient indicates internal consistency that ranges between 0 and 1, with  $\alpha\ge 0.9$ excellent, $0.8\le\alpha<0.9$ good, $0.7\le\alpha<0.8$ acceptable, $0.6\le\alpha<0.7$ questionable, $0.5\le\alpha<0.6$ poor, and $\alpha<0.5$ unacceptable~\cite{gliem2003calculating}. The closer the value to 1, the higher the internal consistency.

For determining internal consistency, we placed all the apps we rated using our proposed rating scale. Table~\ref{Internal consistency} reports the internal consistency of the sub-scales of our devised rating scale -- aesthetics, performance, usability, subjective quality, transparency and perceived impact. We excluded two sub-scales -- general and functionality, as their items are not meant to be collective measures of the construct. The overall internal consistency of our modified scale was high at alpha 0.97, which is excellent~\cite{gliem2003calculating}.

\begin{table}[htbp]
\caption{Internal consistency of the devised app rating scale.}\label{Internal consistency}
\newcolumntype{C}{>{\centering\arraybackslash}X}
\begin{tabularx}{\textwidth}{lCC}
\toprule
App rating sub-scale & Cronbach’s alpha ($\alpha$) & Internal consistency \\
\midrule
Aesthetics & 0.93 & Excellent ($0.9\le\alpha$)\\
Usability & 0.92 & Excellent ($0.9\le\alpha$)\\
Performance & 0.85 & Good ($0.8\le\alpha<0.9$)\\
Subjective & 0.94 & Excellent ($0.9\le\alpha$)\\
Transparency & 0.85 & Good ($0.8\le\alpha<0.9$)\\
Perceived impact & 0.95 & Excellent ($0.9\le\alpha$)\\
\hline
Overall & 0.97 & Excellent ($0.9\le\alpha$) \\
\bottomrule
\end{tabularx}
\end{table}

\subsection{Inter-rater and intra-rater reliability}
Inter-rater reliability is a way of quantifying the level of agreement between two or more raters who rate an item (in this case, an app) independently based on a set of criteria~\cite{lange2011inter}. We used the intra-class correlation (ICC) method to assess inter-rater reliability. ICC is one of the most widely used statistics for evaluating inter-rater reliability if a study includes two or more raters~\cite{sawa2007interrater}. In our study, all apps were rated by the same three raters. Thus, we used the ICC two-way mixed model as it is recommended when the raters are fixed and each of the apps is rated by all raters~\cite{koo2016guideline}. Depending on the 95\% confidence interim of the ICC estimation, values smaller than 0.5, within 0.5 and 0.75, within 0.75 and 0.9, and higher than 0.90 suggest poor, moderate, good, and excellent reliability, respectively~\cite{koo2016guideline}. 
The ICC score of our 17 apps was calculated as 0.869 (95\% CI ranging from 0.847 to 0.888), showing a good level of inter-rater reliability.


Intra-rater reliability is estimated to measure how consistent an individual is at measuring a set of criteria. This is a reliability estimation in which the same evaluation is performed by the same rater on more than one occasion.
To measure the intra-rater reliability of the three raters, the three raters reviewed all 17 apps twice in a two-months interval. All three raters showed a significant good level of intra-rater reliability between their two ratings; their two-way mixed ICC values were 0.921 (95\% CI 0.851 --
0.958), 0.985 (95\% CI 0.972 -- 0.992), and 0.869 (95\% CI 0.847 --
0.888), respectively.

\subsection{Overall assessment of evaluated apps}

The assessment scores for all 17 selected apps are reported in Table \ref{assesment-table}, 
with sub-scale ratings and their mean, standard deviation. The mean of the scores of all criteria in a sub-scale is used to calculate the rating of that sub-scale. The overall mean and standard deviation of an app's scores in all sub-scales were used to evaluate its overall quality.

\begin{table}[ht]
\centering
\caption{Assessment scores for plant disease detection apps.}
\label{assesment-table}
		\newcolumntype{C}{>{\centering\arraybackslash}X}
\resizebox{1\textwidth}{!}{
\begin{tabular}{l l l l l l l l l c}
\toprule
App name 
& \makecell[t l]{Aesth\\etics} 
& \makecell[t l]{Gene\\ral} 
& \makecell[t l]{Perfor\\mance} 
& \makecell[t l]{Usabi\\lity}
& \makecell[t l]{Functio\\nality} 
& \makecell[t l]{Subje\\ctive} 
& \makecell[t l]{Transp\\arency} 
& Impact
& \makecell[t c]{Mean\\(Std dev)} 
\\
\midrule

Agrio & 4.75 & 2.25 & 4.67 & 4.00 & 3.57 & 2.83 &2.80 & 3.83 & 3.53 (1.56)\\

\makecell[t l]{AgroAI - Plant Diseases Diagnosis (Early Access)} &4.75 &2.13 &4.83 & 3.67 &3.00 &4.83 &3.80 &4.15 &3.81 (1.50)\\

\makecell[t l]{Cassava Plant Disease Identify} & 2.25 & 3.25 & 4.33 & 1.83 & 1.67 & 1.33 & 2.20 & 3.00 & 2.48 (0.99)\\ 

Cropalyser & 5.00 & 3.13 & 4.83 & 3.17 & 2.14 & 3.33 & 4.40 & 3.83 & 3.65 (1.41)\\

\makecell[t l]{Garden Plants Diseases Detector} & 3.00 & 3.25 & 4.33 & 2.83 & 2.71 & 2.50 & 3.40 & 3.17 & 3.19 (1.47)\\

\makecell[t l]{Leaf Doctor} & 5.00 & 3.50 & 4.83 & 3.00 & 1.86 & 3.00 & 4.40 & 1.17 & 3.33 (1.77)\\

Leafy &	3.75 & 2.38 &  5.00 & 4.00 & 3.29 & 3.00 & 2.80 & 2.33 & 3.26 (1.69)\\

\makecell[t l]{PDDApp: plant disease detection}	& 3.00 & 2.38 & 3.50 &	2.83 & 2.43 &2.17 & 2.80 & 1.00 & 2.44 (1.55)\\

\makecell[t l]{Pestoz Identify Plant diseases} & 4.75 & 2.38 & 4.67 & 5.00 & 2.71 & 4.00 & 3.40 &  3.33 &  3.56 (1.48)\\

\makecell[t l]{Plant Diseases and Pests} & 1.00 & 2.75 & 4.50 & 1.50 & 3.00 & 1.83 & 2.60 & 1.67 & 2.42 (1.69)\\

\makecell[t l]{Plant Disease Detector} & 4.75 & 2.50 & 3.33 & 3.00 & 2.71 & 1.67 & 1.80 & 1.00 & 2.63 (1.75)\\

\makecell[t l]{Plant Disease Identifier} &3.00 &3.25 &4.33 &2.67 &2.71 &2.67 &3.40 &3.17 & 3.19 (1.47)\\
\makecell[t l]{PlantDoctor}	& 2.50 &2.00 &3.00  & 1.83 &1.57 &1.00 &1.00 &1.00 & 1.81 (1.35)\\

\makecell[t l]{PlantifyDr}\ &3.50 & 2.00 &4.83 & 4.17 &3.00 & 3.67 &2.80 & 3.67  & 3.37 (1.65)\\

Plantix - your crop doctor &5.00	&4.13 &5.00 &4.33 &3.29 &5.00 &5.00 &5.00 &4.56 (1.16)\\

\makecell[t l]{Plants Disease Identification} & 2.00 & 2.50 & 4.50 & 3.50 & 2.71 & 1.83 & 2.80 & 3.00 & 3.00 (1.53)\\

Riceye & 3.00 & 2.25 & 4.83 & 2.50 & 2.43 & 1.83 & 2.20 & 1.00 & 2.53 (1.65)\\
\bottomrule
\end{tabular}
}
\end{table}

In aesthetics, 41.18\% (7/17) of the apps are rated above 4 out of 5. Among them three apps, \app{Cropalyser}, \app{Plantix - your crop doctor}, and \app{Leaf Doctor}, receive the highest score, i.e., 5 out of 5. Only one app, \app{Plant Diseases and Pests}, receives the lowest score of 1.

In general characteristics, all the apps except \app{Plantix - your crop doctor}, received a score below 4. In terms of login or sign-up, 82.36\% (14/17) of the apps do not require users to login or sign-up to use - this allows the rural farmers and gardeners to easily benefit from the app. However, only three apps (17.67\%) send users notifications about their regular tasks. Most of the apps 94.11\% (16/17) do not have a premium subscription. The user cannot export their data in 70.59\% (12/17) of the apps. 52.94\% (9/17) of the apps do not present information visually through charts, graphs, images or videos and do not have any tutorial or on-boarding facilities to help users operate the apps. Also, 64.71\% (11/80) of the apps do not have customization features. In this sub-scale, the app \app{Plantix - your crop doctor} received the highest score of 4.13. 

In the performance sub-scale, 14 out of 17 apps (82.36\%) received a high score of between 4.33 and 5, which means they are responsive, their components are working well, the apps do not crash, they consume low battery power and have a reasonable memory. Among those 14 apps, \app{Leafy} and \app{Plantix - your crop doctor}, scored 5 out of 5, which indicates that these apps show excellent performance while being used.

In the usability sub-scale, 82.36\% (14/17) of the apps were very easy to use and scored 4 to 5, 70.59\% (12/17) had high navigational accuracy, and 58.82\% (10/17) depict a high quality of gestural design. Overall, only a few apps 29.41\% (5/17) scored high (between 4 and 5) in this sub-scale. 
The lowest rated app in this sub-scale is \app{Plant Diseases and Pests}, which scored 1.50. Only the app \app{Pestoz - Identify Plant diseases} scored 5 out of 5 in usability.

The functionality sub-scale is the most significant sub-scale as it evaluates the features an app provides for disease identification and management. In our review, only 6 out 17 apps (35.29\%) scored between 3 and 3.57 and the rest scored below 3. The results indicate that apps with proper automation of identifying plants and diseases along with disease severity estimation are mostly missing in the current apps.  

The transparency sub-scale is assessed based on an app’s description in the app store, credibility through a legitimate source, verified by evidence, goals and policy in accessing and sharing user data. In this sub-scale, only 3 out of 17 apps (17.65\%) scored above 4. Among them, \app{Plantix - your crop doctor} scored 5 out of 5. Two apps, \app{Cropalyser} and \app{Leaf Doctor}, each scored 4.40.

The subjective quality sub-scale is measured based on an individual’s willingness to use, recommend people and pay for the apps, and the overall star rating given by raters. In this sub-scale, 3 out of 17 apps (17.65\%) scored 4 and higher. Among them, \app{Plantix - your crop doctor} scored the highest value of 5. \app{PlantDoctor} scored the lowest value of 1.

The perceived impact on user sub-scale measures the effectiveness of an app in changing the attitudes of the farmers and plant lovers who use the app. This was evaluated based on whether the app provides a community or forum to share info and seek help. Eight out of 17 apps (47.05\%) help enormously to increase awareness of the importance of addressing plant diseases and behavior change toward crop illness treatment, 58.82\% (10/17) are likely to increase knowledge or understanding of infected plants; 29.41\% (5/17) do not have concern about these observations. Two out of 17 apps (11.76\%) (\app{AgroAI - Plant Diseases Diagnosis} and \app{Plantix - your crop doctor}) rated 4.15 and 5, respectively. They can change attitudes toward improving plantations and care. The app, \app{Plantix - your crop doctor}, provides strong motivation for users in their gardening and farming. On the other hand, 23.53\% (4/17) of the apps scored 1 in this sub-scale.

From the overall screening of the mean values of the evaluated apps, \app{Plantix - your crop doctor}, scored    the highest (4.56), but the functionality score for this app was 3.29. Most of the apps (10/17) got mean values ranging from 3 to 4, indicating that the quality of the apps is not up to the mark and they are specifically lagging in functionality.

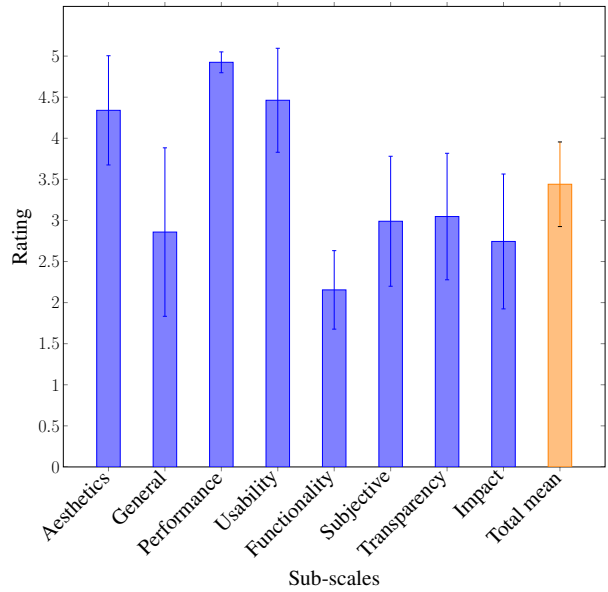
\begin{figure}[ht]
\centering
\resizebox{8cm}{!}{
\begin{tikzpicture}
    \begin{axis}[
        symbolic x coords={Aesthetics, General, Performance, Usability, Functionality,  Subjective, Transparency, Impact, Total mean},
        x tick label style={font=\huge, rotate=45, anchor=east},
        ybar=-0.8cm,
        ymin=0,
        xtick distance=1,
        bar width=0.8cm,
        ytick={0,0.5,1,1.5,2,2.5,3,3.5,4,4.5,5},
        ylabel={Rating},  
        xlabel={Sub-scales},
        label style={font=\huge},
        y tick label style={font=\LARGE},
    ]
        \addplot+ [draw = blue,
       fill=blue!50,
            error bars/.cd,
                y dir=both,
                y explicit,
        ] coordinates {
            (Aesthetics,4.34) +-(0,0.6647)
            (General,2.8585) +- (0,1.0253)
            (Performance,4.9246) +- (0,0.1268)
            (Usability,4.4625) +- (0,0.6328)
            (Functionality,2.1547) +- (0,0.4777)
            (Subjective,2.99) +- (0,0.791)
            (Transparency,3.0475) +- (0,0.7698)
            (Impact,2.7443) +- (0,0.8206)
        };
        \addplot [draw =orange,
        fill = orange!50,
            error bars/.cd,
                y dir=both,
                y explicit,
        ] coordinates {
        (Total mean,3.4404) +- (0,0.5149)
        };
    \end{axis}
\end{tikzpicture}
}
\caption{Sub-scale specific ratings and overall rating.}  
   \label{overall-app-rating}   
\end{figure}

Fig.~\ref{overall-app-rating} shows the average sub-scale specific scores and the total mean score of all 17 apps. The overall performance and usability of the evaluated apps are very good (scored above 4.5), which means they are easy to use and fast to open. However, the overall aesthetics value is average because few apps are outstanding and eye-catchy to look at, and most are not so attractive. Our focus is the functionality sub-scale, which scored below 2.5. This value indicates that we have not found any quality apps that can do all the desired tasks regarding plant disease identification.

\subsection{Assessment of app functionality}

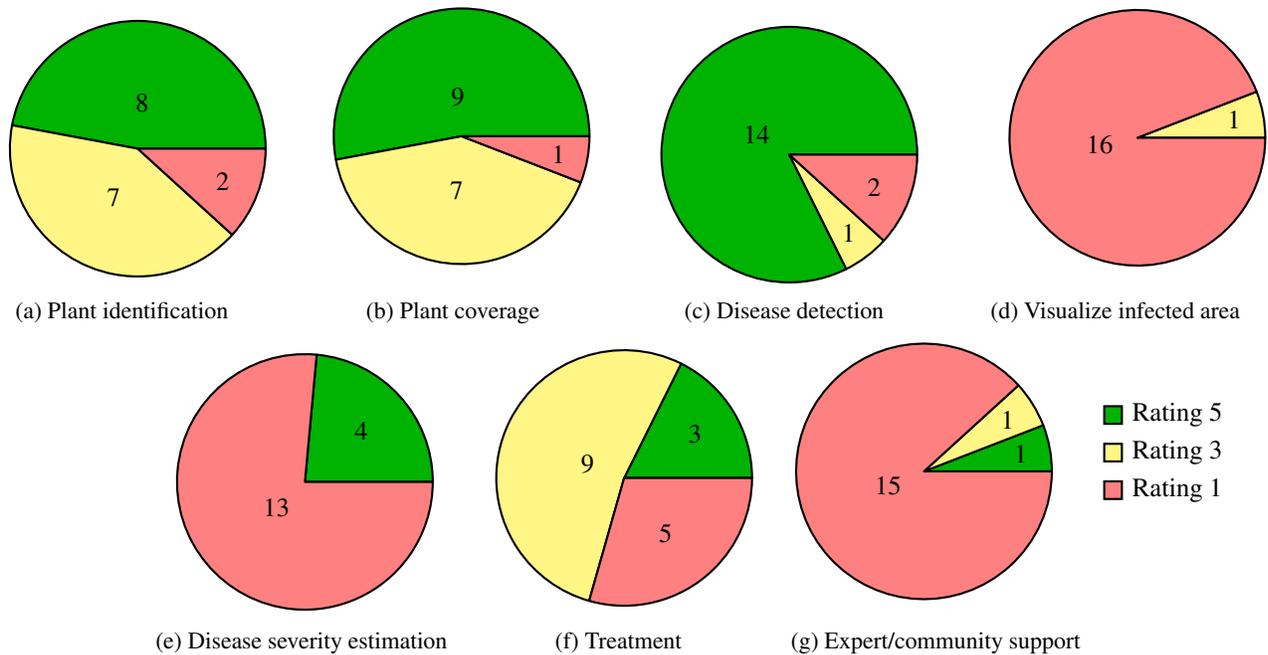
\begin{figure}[htbp]
\centering
\begin{subfigure}[b]{0.2\textwidth}
\centering
\begin{tikzpicture}
  \pie[sum=auto, radius=1.7,color={green!70!black, yellow!60, red!50}]{8/, 7/, 2/}
  \end{tikzpicture}
  \caption{Plant identification}
\end{subfigure}
\hfill
\begin{subfigure}[b]{0.2\textwidth}
  \centering
  \begin{tikzpicture}
  \pie[sum=auto,  radius=1.7,color={green!70!black, yellow!60, red!50}]{9/,  7/, 1/}
  \end{tikzpicture}
  \caption{Plant coverage}
\end{subfigure}
\hfill
\begin{subfigure}[b]{0.2\textwidth}
  \centering
  \begin{tikzpicture}
  \pie[sum=auto,  radius=1.7,color={green!70!black, yellow!60, red!50}]{14/,  1/, 2/}
  \end{tikzpicture}
  \caption{Disease detection}
\end{subfigure}
\hfill
\begin{subfigure}[b]{0.2\textwidth}
  \centering
  \begin{tikzpicture}
  \pie[sum=auto,  radius=1.7,color={yellow!60, red!50}]{1/, 16/}
  \end{tikzpicture}
  
  \caption{Visualize infected area}
\end{subfigure}
\hfill
\begin{subfigure}[b]{0.25\textwidth}
  \centering
  \begin{tikzpicture}
  \pie[sum=auto,  radius=1.7,color={green!70!black, red!50}]{4/, 13/}
  \end{tikzpicture}
  \caption{Disease severity estimation}
\end{subfigure}
\begin{subfigure}[b]{0.25\textwidth}
  \centering
  \begin{tikzpicture}
  \pie[sum=auto,  radius=1.7,color={green!70!black, yellow!60, red!50}]{3/,  9/, 5/}
  \end{tikzpicture}
  \caption{Treatment}
\end{subfigure}
\begin{subfigure}[b]{0.25\textwidth}
  \centering
  \begin{tikzpicture}
  \pie[text=legend, sum=auto,  radius=1.7,color={green!70!black, yellow!60, red!50}]{1/Rating 5,  1/Rating 3, 15/Rating 1}
  \end{tikzpicture}
  \caption{Expert/community support}
\end{subfigure}
%
\caption{Representation of specific functionality rating ratio.}
\label{ratio pi chart}
\end{figure}

Plant disease identification apps should contain some major criteria for fulfilling their sole intention. These apps are meant to help farmers and gardeners by providing solutions regarding their plants. However, the reality is different because the apps lack the functionalities to provide necessary to help the target users. We have already discussed seven critical characteristics for plant disease detection apps in Section~\ref{Func}. They are (1) plant identification; (2) plant coverage; (3) disease detection; (4) visualize infected area; (5) disease severity estimation; (6) treatment; and (7) expert/community support. The outcomes of our functionality analysis of 17 evaluated applications based on the seven assessment criteria are shown in Fig.~\ref{ratio pi chart}.

\begin{figure}[htbp]
\centering
    \includegraphics[width=.7\textwidth]{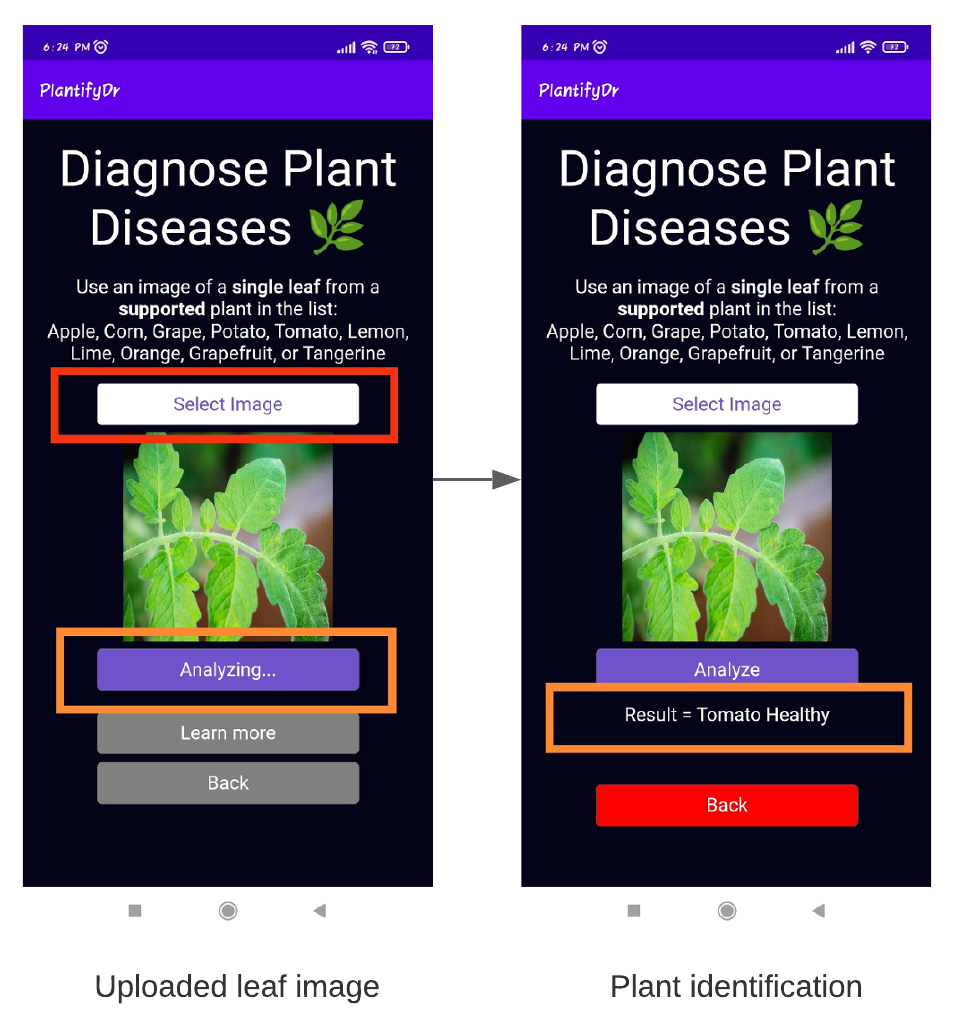}
  \caption{Plant identification by \app{PlantifyDr}.}
  \label{Plantidenti}
\end{figure}

First, identifying the plants is a preliminary criterion for plant disease detection apps. Among the 17 reviewed apps, 47.06\% (8/17) can automatically identify plants from the given image and 41.18\% (7/17) kept the option of choosing the plants manually before diagnosing any disease. Only two apps, \app{Leaf Doctor} and \app{Riceye}, do not fulfill this criteria as \app{Leaf Doctor} focuses only on disease severity and visualizing the infected area of the leaf, where \app{Riceye} is dedicated for rice crops only. Different apps have adopted different technologies to identify plants. For example, the \app{PlantifyDr} app uses ML algorithms to identify a plant and detect whether a specific plant has a disease or not. Plant identification by \app{PlantifyDr} is shown in the Fig.~\ref{Plantidenti}. Another key criterion in our study was to identify the plant coverage, i.e., the number of plants covered by the disease detection apps. The only app that missed this criterion is \app{Leaf Doctor}. Most of the apps, nearly 53\% (9/17), can identify diseases of more than 10 plants.

\begin{figure}[htbp]
\centering
  \begin{subfigure}[b]{.49\textwidth}
    \includegraphics[width=\textwidth]{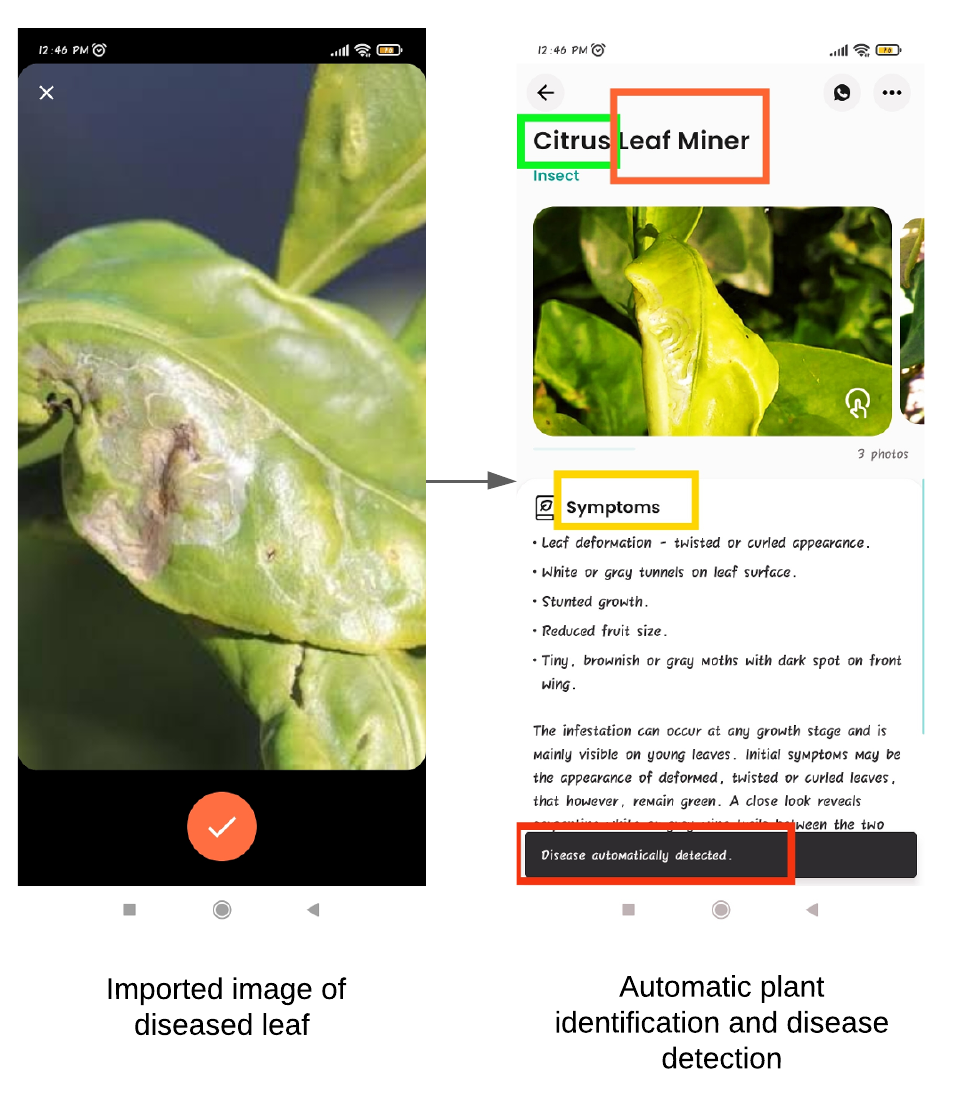}
    \caption{\app{Plantix - your crop doctor}}
    \label{PlantixD}
  \end{subfigure}
  \hfill
  \begin{subfigure}[b]{0.49\textwidth}
    \includegraphics[width=\textwidth]{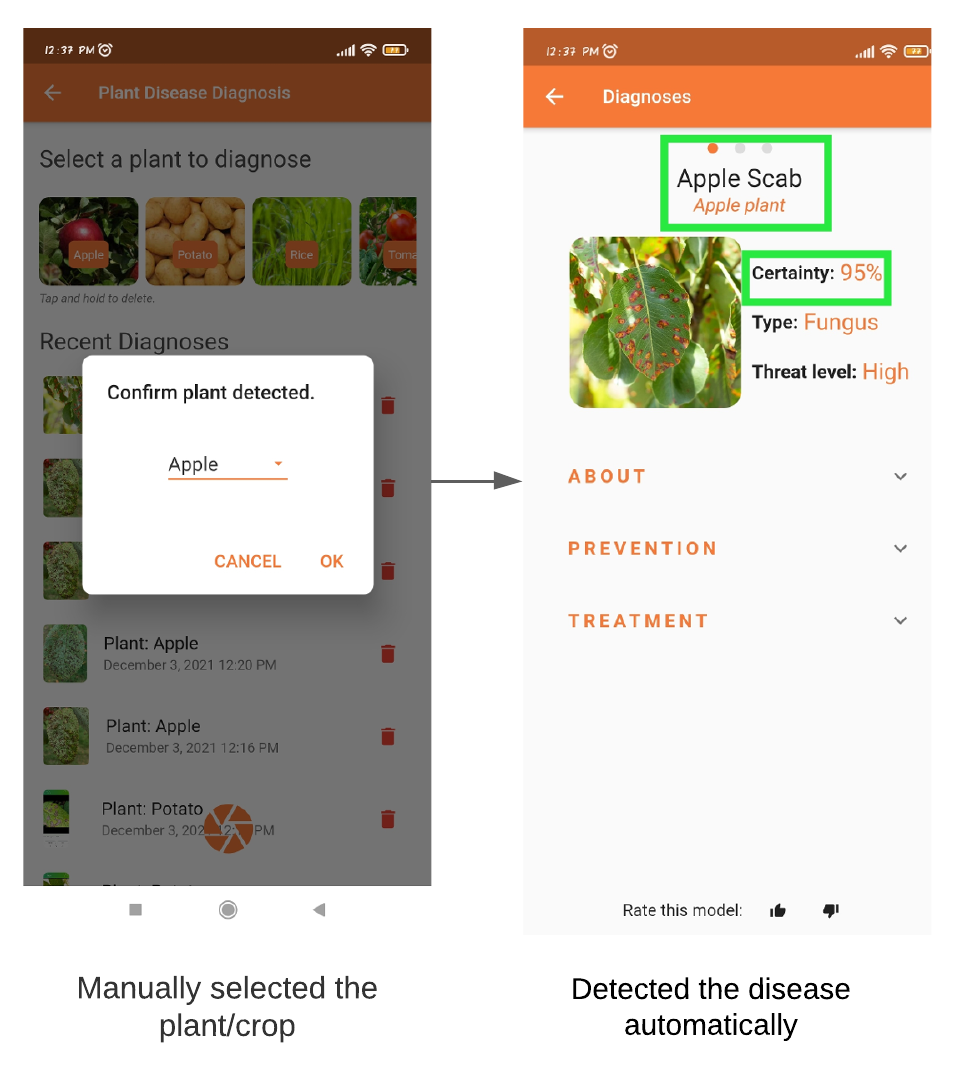}
    \caption{\app{AgroAI - Plant Diseases Diagnosis (Early Access)}}
    \label{Agroai}
  \end{subfigure}
  \caption{Disease detection.}
  \label{Disease Detection}
\end{figure}

Our most significant functionality is disease detection where it is expected that an app would automatically recognize the disease from the photos of affected plants or leaves. Fortunately, 82\% (14/17) of the apps have fulfilled that expectation. Only one app, \app{Cropalyser}, identifies diseases  based on answers given by the user from a series of questionnaires. On the other hand, the \app{Leaf Doctor} and \app{Plant Doctor} apps do not provide this functionality. Our reviewed apps use a variety of techniques to detect diseases. For instance, \app{Plants Disease Identification} app uses ML Apple technology to classify the plant diseases productively. The apps \app{Cassava Plant Disease Identify} and \app{Garden Plant Diseases Detector} use computer vision techniques to classify the disease and monitor severity. To get the maximum accuracy in disease detection, a large image database containing thousands of images has been used for developing a model using artificial neural networks (ANN). The app \app{Cassava Plant Disease Identify} allows user to report photos that are not properly captured. As more photographs are uploaded, the accuracy of the database improves. User can notify the developer of the app by email, if a photograph is not recognized. The app is improving its accuracy over time by using ML. The app \app{Pestoz- Identify Plant diseases} also applies advanced computer vision techniques to identify the diseases. Fig.~\ref{Disease Detection} shows screenshots of the disease detection UI of the \app{Plantix} and \app{AgroAI} apps.

Another important criterion is disease severity estimation which is provided by only four apps, \app{AgroAI - Plant Diseases Diagnosis (Early Access)}, \app{Leaf Doctor}, \app{Riceye}, and \app{Cassava Plant Disease Identify}. Visualization is also a useful measure to assess the severity of disease and identify the infected area of a plant leaf. 

\begin{figure}[!htbp]
\centering
    \includegraphics[width=\textwidth]{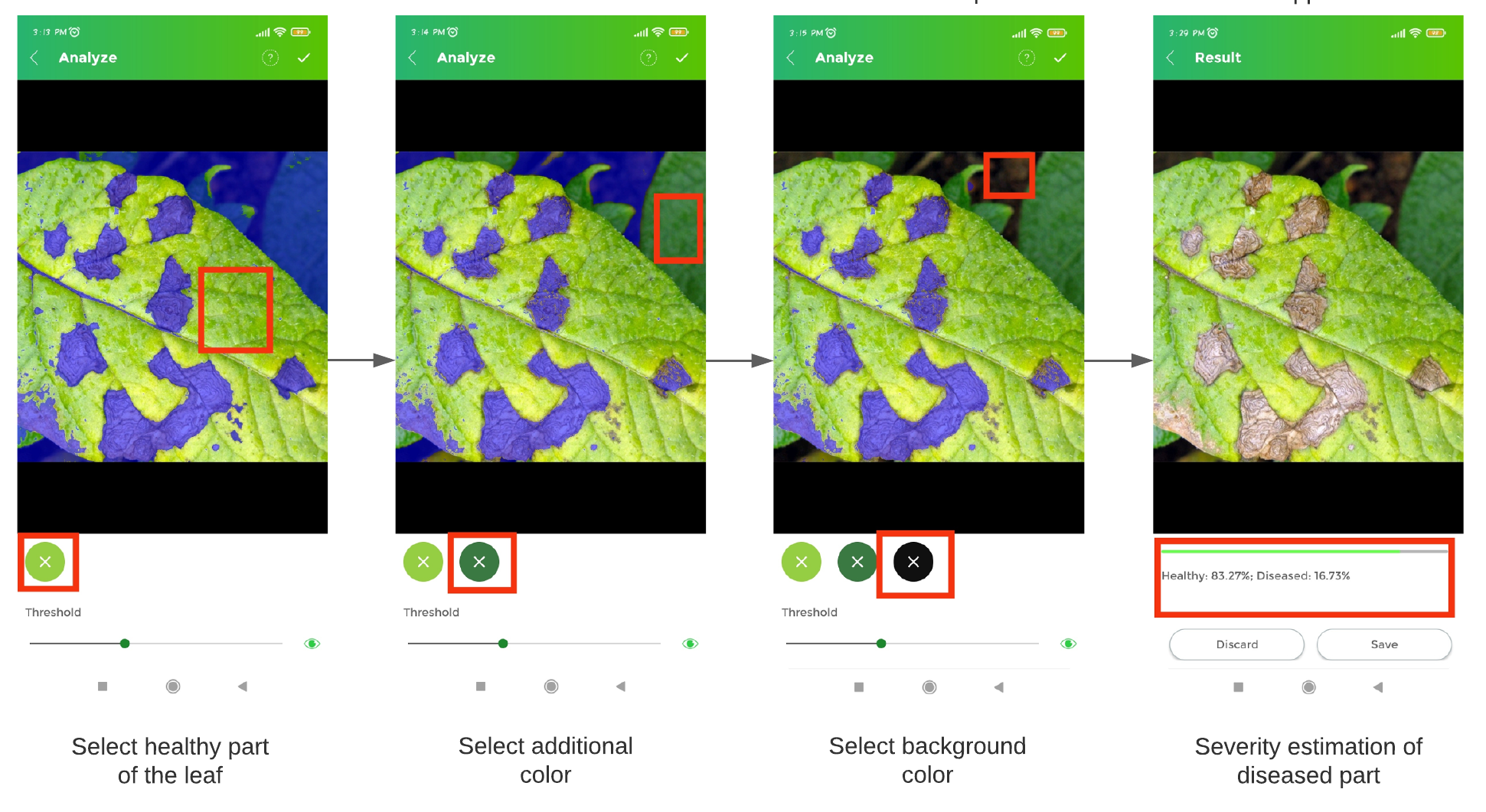}
  \caption{Severity detection and visualizing the infected area of diseased leaf by \app{Leaf Doctor}.}
  \label{LDseverity}
\end{figure}

Only \app{Leaf Doctor} provides this functionality. This app helps users differentiate between damaged and healthy leaf tissues and calculates disease severity as a percentage, as shown in Fig.~\ref{LDseverity}. However, the app requires user knowledge and input. The user needs to select up to 8 distinct colors that indicate healthy tissues on the app's touch panel and subsequently moves a threshold slider until only symptomatic tissues are changed into a bluish shade. Afterwards, the app calculates the diseased percentage using that pixelated photo. The precision, accuracy, and stability of the \app{Leaf Doctor} app have been tested using six diseases and typical lesions of increasing severity~\cite{pethybridge2015leaf}. 

\begin{figure}[!htbp]
\centering
  \begin{subfigure}[b]{0.49\textwidth}
    \includegraphics[width=\textwidth]{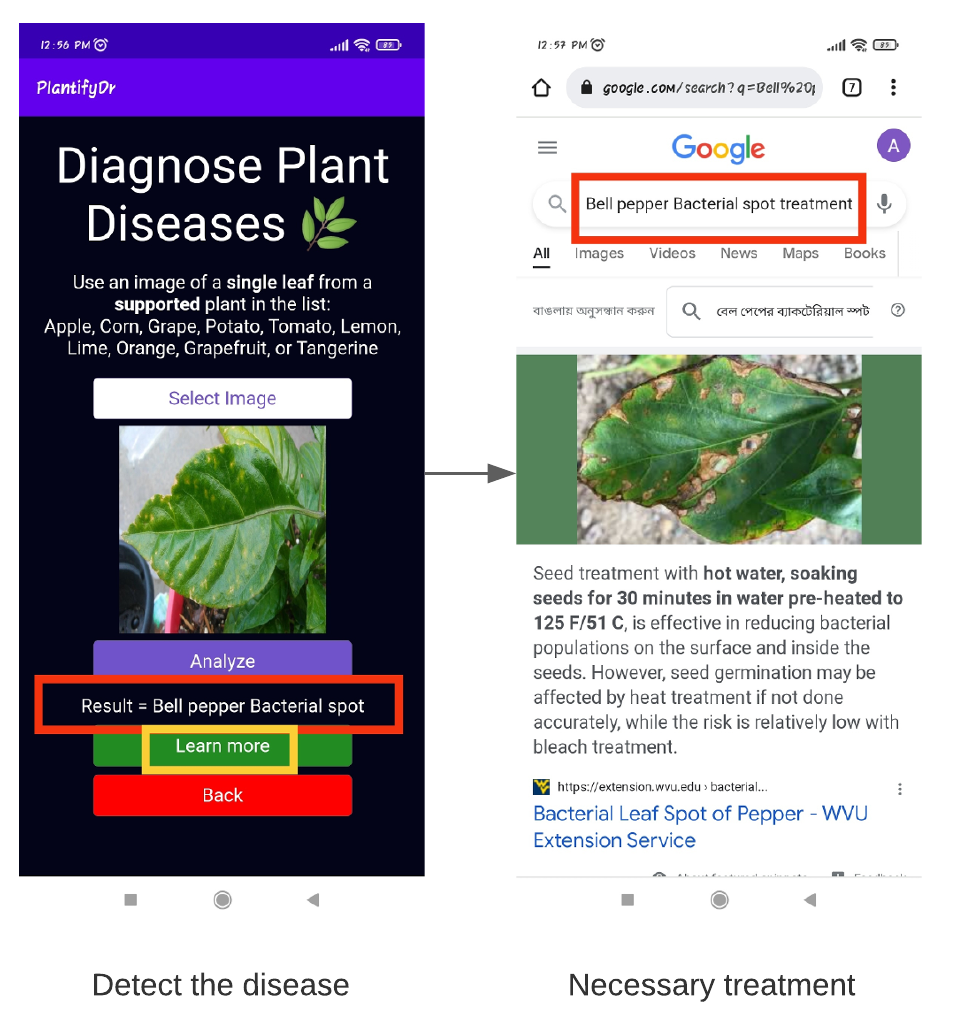}
    \caption{\app{PlantifyDr}}
    \label{PlantifyDrT}
  \end{subfigure}
  \hfill
  \begin{subfigure}[b]{.49\textwidth}
    \includegraphics[width=\textwidth]{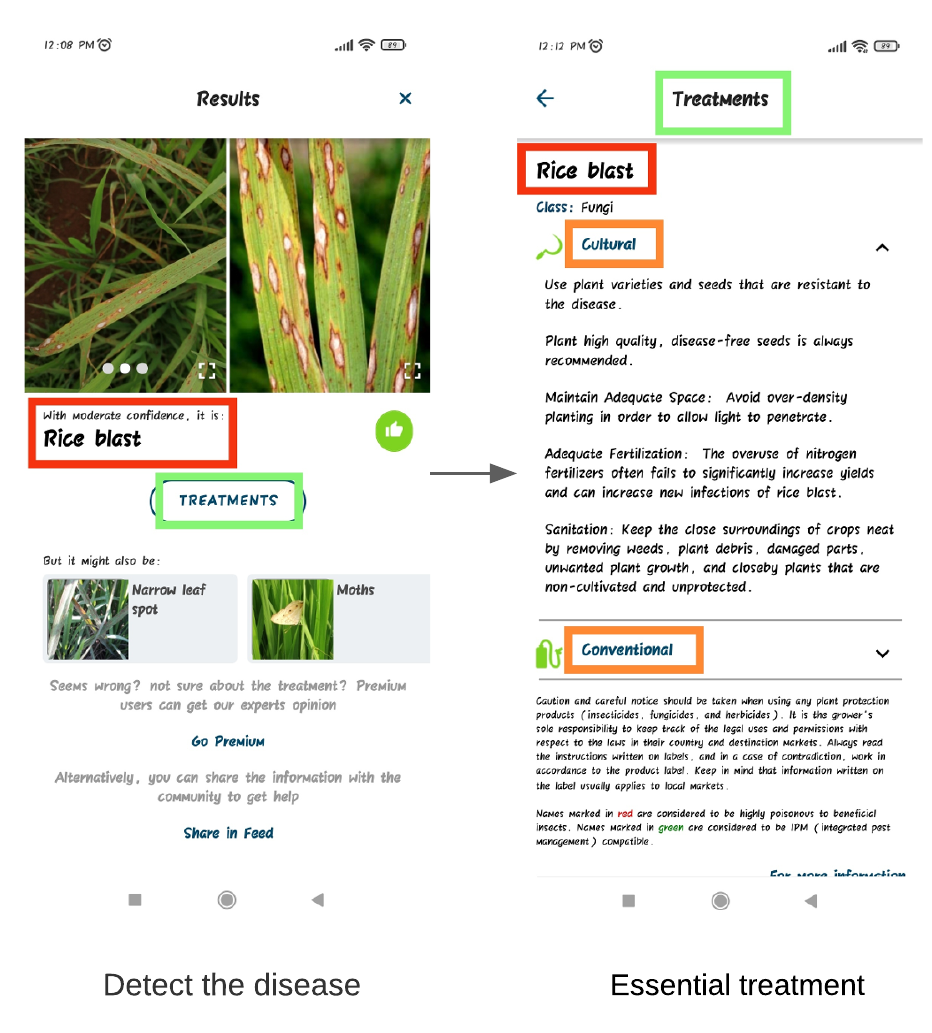}
    \caption{\app{Agrio}}
    \label{AgrioT}
  \end{subfigure}
  \caption{Treatment.}
  \label{Treatment}
\end{figure}

Providing treatment and care suggestions for plants is an essential functionality to motivate farmers and gardeners to use plant disease detection apps. Two apps, \app{Leafy} and \app{PlantifyDr}, detect the plant disease and subsequently take users directly to Google to get the latest information about the treatment (see Fig.~\ref{PlantifyDrT} for UI screenshots of \app{PlantifyDr}). Nine apps provide treatment information from their own database. Only the app \app{Agrio} provides treatment through a community (see Fig.~\ref{AgrioT}).

Local farmers can contact with field investigators through \app{Agrio} app to receive benefits. This app uses drones for remote field investigation. Users of \app{Agrio} app can look at remote sensing insights, alerts, and data gathered during the inspection process. They have a community where farmers can share their problems and intelligent sources quickly provide necessary solutions or treatments. Within the groups, discussions about the necessary interventions are becoming simpler. This app also uses AI which makes precise hyper-local weather forecasts easily available to all growers. Growing degree days (GDD) is a measure used to calculate the amount of heat required for the development of organisms (such as insects) in each stage of their growth. Monitoring GDD helps eliminate the guesswork in determining the time required for control measures. One of the exciting possibilities is the alignment of treatment schedules in different farms and gardens.  Area-wide integrated pest management is the paradigm in which pest control decisions and timing are coordinated across many fields.

\app{Agrio}'s daily briefing informs growers of the required scouting operations and interventions in their fields and also uses big-data to optimize predictions and offer phenology models specific to the different locations. They coordinate area-wide integrated pest management operations and present users with the optimal treatment on time using AI. Growers can use image identification capabilities if help is needed with the trap analysis.

\begin{figure}[htb]
\centering
  \begin{subfigure}[b]{.45\textwidth}
    \includegraphics[width=\textwidth]{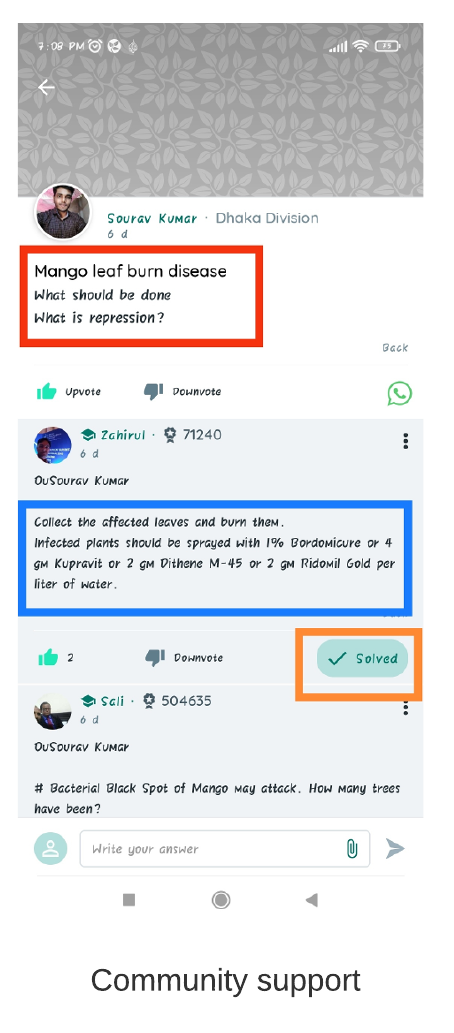}
    \caption{\app{Plantix - your crop doctor}}
    \label{PlantixC}
  \end{subfigure}
  \begin{subfigure}[b]{0.45\textwidth}
    \includegraphics[width=\textwidth]{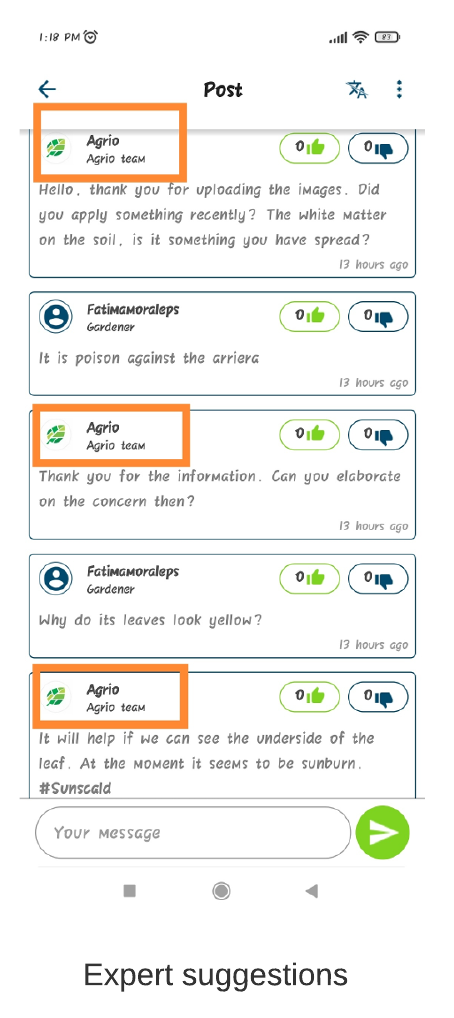}
    \caption{\app{Agrio}}
    \label{AgrioEx}
  \end{subfigure}
  \caption{Expert/community support.}
  \label{ExpertCommunity}
\end{figure}

One extraordinary feature to connect the user with fellow gardeners, farmers and experts is the creation of a community. Only two apps provide this feature, \app{Plantix - your crop doctor} and \app{Agrio}. Fig.~\ref{ExpertCommunity} shows the community support UI screenshots of these two apps.

The \app{Plantix - your crop doctor} community is the world's largest online community of farmers and agricultural specialists where users can take advantage of agricultural professionals' knowledge and share their crop related problems to get solutions. The \app{Agrio} app provides a community with experts who respond to the users' problems more individually and provide treatment based on the issue, as discussed above. 
In conclusion, this expert/community support feature should be included in all the apps for better user connectivity.

Our overall assessment of the functionality criteria for the apps is illustrated in Table \ref{table5}. Here, we can see that 88.23\% of the apps can identify plants and detect diseases. In these sectors, apps from both app stores showed almost equal contributions. The most important criterion for our apps was disease detection but two of the apps \app{Leaf Doctor} and \app{PlantDoctor} did not provide this functionality. About 94\% of the apps cover more than 10 plants, and, in this case, Android apps have a richer plant disease database than iOS apps. Sixty four percent of the apps provide treatment after diagnosing plant disease. Most of our reviewed apps lack three functionalities - disease severity estimation, visualizing the infected area, and expert/community support. Of all the apps, 23.52\% can estimate the severity of the disease from a leaf, 11.76\% have expert/community support, and only one app can visualize the infected area from an infected leaf.

Among the 17 reviewed apps, \app{Agrio}, \app{AgroAI - Plant Diseases Diagnosis}, \app{Cassava Plant Disease Identify} and \app{Plantix - your crop doctor}, contain a total of five out of seven functionalities. Seven apps, \app{Pestoz Identify Plant diseases}, \app{Cropalyser}, \app{PDDApp: plant disease detection} , \app{Leafy}, \app{Plant Disease Detector},\app{PlantifyDr}, and \app{Plant Diseases and Pests}, contain four functionalities. The remaining six apps contained less than half the functionalities outlined. In our study, no app was found to have all seven functionalities for the best plant disease detection and management.

\begin{table}[htbp]
\centering
\caption{Assessment criteria for the functionality of apps.}\label{table5}
\newcolumntype{C}{>{\centering\arraybackslash}X}
\begin{tabularx}{\textwidth}{lCCC}
\toprule
 Functionality assessment criteria & \makecell[t c]{Google Play (n=9)\\n (\%)} & \makecell[t c]{Apple App (n=8)\\n (\%)} & \makecell[t c]{Total (N=17)\\N (\%)}\\
\midrule
Plant identification &  8 (88.89) &  7 (87.50) & 15 (88.23)\\
Plant coverage & 9 (100) & 7 (87.50) & 16 (94.12)\\
Disease detection & 8 (88.89) & 7 (87.50) &  15 (88.23)\\
Disease severity estimation & 2 (22.22) & 2 (25) &  4 (23.52)\\
Visualize infected area & 0 (0) & 1 (12.50) &  1 (5.88)\\
Treatment &  7 (77.78) &  4 (50) & 11 (64.71)\\
Expert/Community support &  1 (11.11) &  1 (12.50) &  2 (11.76)\\
\bottomrule
\end{tabularx}
\end{table}

\subsection{Analysis of app ratings from the app store and the developed rating scale}

\begin{figure}[!htb]
\centering
\resizebox{.8\textwidth}{!}{
\begin{tikzpicture}
\begin{axis}[
    xbar,
    bar width= 9pt,
    ylabel={App Name},  
    xlabel={App rating/score},
   label style={font=\LARGE},
    legend style={at={(0.5,-0.1),font=\LARGE},
      anchor=north,legend columns=-1},
    symbolic y coords={PDDApp: plant disease detection,Plants Disease Identification,PlantDoctor,Leaf Doctor,Plant Disease Identifier,Pestoz Identify Plant diseases,Agrio,Cropalyser,Plantix - your crop doctor,PlatifyDr},
    ytick=data,
    xtick={0,1,2,3,4,5},
    xmin=0,
    xmax=5.1,
    enlarge y limits  = 0.05,
    enlarge x limits  = 0.01,
    scaled ticks=false,
    x tick label style={font=\LARGE},
    y tick label style={font=\LARGE},
    ytick distance=1,
    ]
 
 \addplot coordinates {(2.44,PDDApp: plant disease detection) (3,Plants Disease Identification) (1.81,PlantDoctor) (3.33,Leaf Doctor) (3.19,Plant Disease Identifier) (3.56,Pestoz Identify Plant diseases) (3.81,Agrio) (3.65,Cropalyser) (4.56,Plantix - your crop doctor) (3.37,PlatifyDr)};

\addplot coordinates {(1,PDDApp: plant disease detection) (1,Plants Disease Identification) (2,PlantDoctor) (2.5,Leaf Doctor)(3.7,Plant Disease Identifier) (3.6,Pestoz Identify Plant diseases) (4,Agrio) (4.5,Cropalyser) (4.3,Plantix - your crop doctor) (5,PlatifyDr)};

\legend{ Measured score, App store rating}
\end{axis}
\end{tikzpicture}}
\caption{App store ratings (omitted apps those ratings were not available) and measured ratings of the apps.}
\label{fig:app-ratings}
\end{figure}
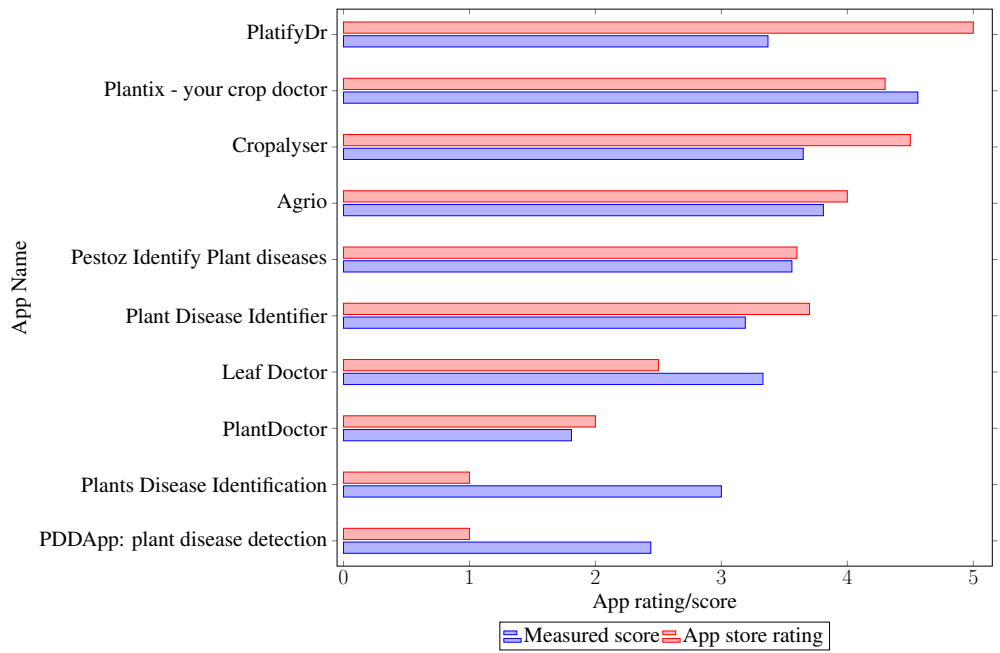

The app stores' ratings and our rating scale's ratings of the evaluated apps were analyzed. Reviews and ratings help users decide whether to download an app. According to a study~\cite{6636712}, mobile consumers do not even bother downloading an app with a rating of fewer than three stars. Additionally, 79\% of users will check at least one review before downloading the application. Good ratings indicate that the app provides an advantage to its users. Developers can do everything to improve the app's rating and make it more successful in the future. During this survey period, \app{AgroAI - Plant Diseases Diagnosis (Early Access)}, \app{Leafy}, \app{Plant Disease Detector}, \app{Riceye}, \app{Cassava Plant Disease Identify}, \app{Garden Plants Diseases Detector}, \app{Plant Diseases and Pests} and \app{Agrio} apps dis not have any star ratings. Thus, these seven apps were excluded from this analysis. Fig.~\ref{fig:app-ratings} shows the app store rating and our measured app rating of ten apps. 

Three apps (\app{PlatifyDr}, \app{Cropalyser}, and \app{Plant Disease Identifier}) got a higher star rating from app store users, but some limitations were found in our analysis, they did not perform well in our devised rating scale. Consequently, these apps achieved a lower rating from our raters. The measured score and app store ratings are almost the same for four apps (\app{Agrio}, \app{Pestoz Identify Plant diseases}, \app{PlantDoctor}, and \app{Plantix - your crop doctor}), which shows that our raters agree with the users' ratings found from the app stores. Finally, three apps (\app{Leaf Doctor}, \app{Plants Disease Identification}, and \app{PDDApp: plant disease detection}) received a slightly higher score compared to the app stores' ratings.


\subsection{Analysis of user reviews from the app stores}
Text reviews are more reliable than star ratings in describing users' actual experiences.
After using a mobile app, users can add comments in the user reviews about that app. The user review section of an app store represents the experience of using the app. Users' feedback helps
potential new users and developers to understand an app's quality. The developer benefits from the evaluation because it focuses on the app's limitations and technical flaws, allowing them to create a better version of the app. 

\begin{figure}[!htbp]
\centering
  \begin{subfigure}[b]{.49\textwidth}
    \includegraphics[width=\textwidth]{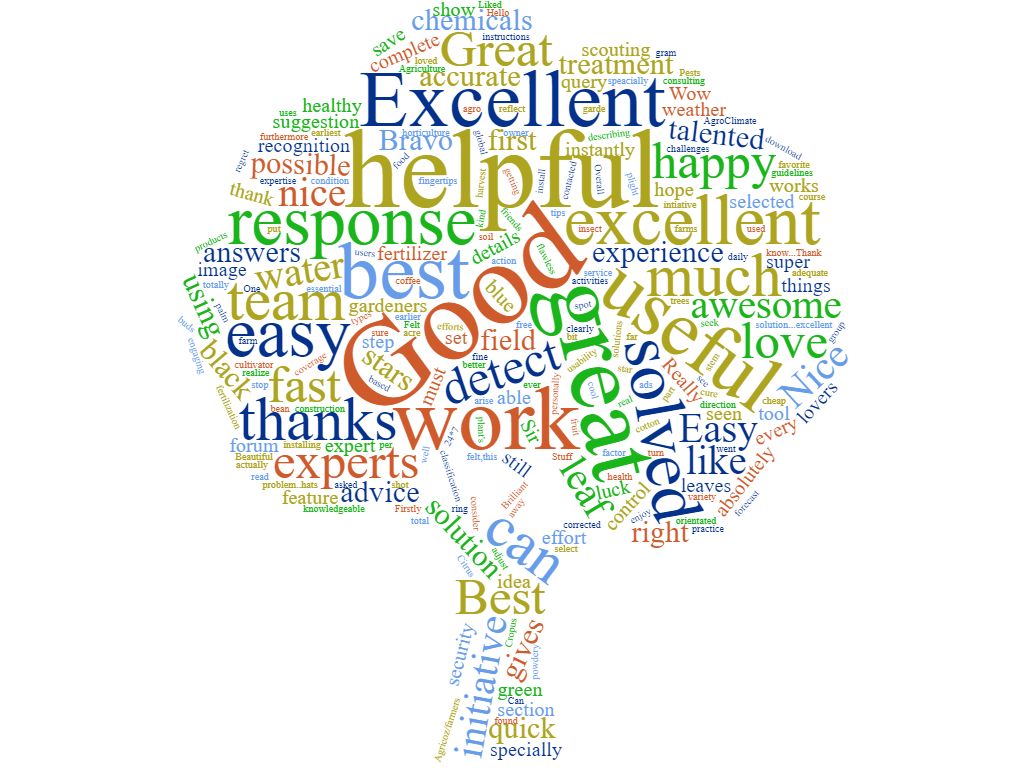}
    \caption{Positive comments}
    \label{good}
  \end{subfigure}
  \begin{subfigure}[b]{0.37\textwidth}
    \includegraphics[width=\textwidth]{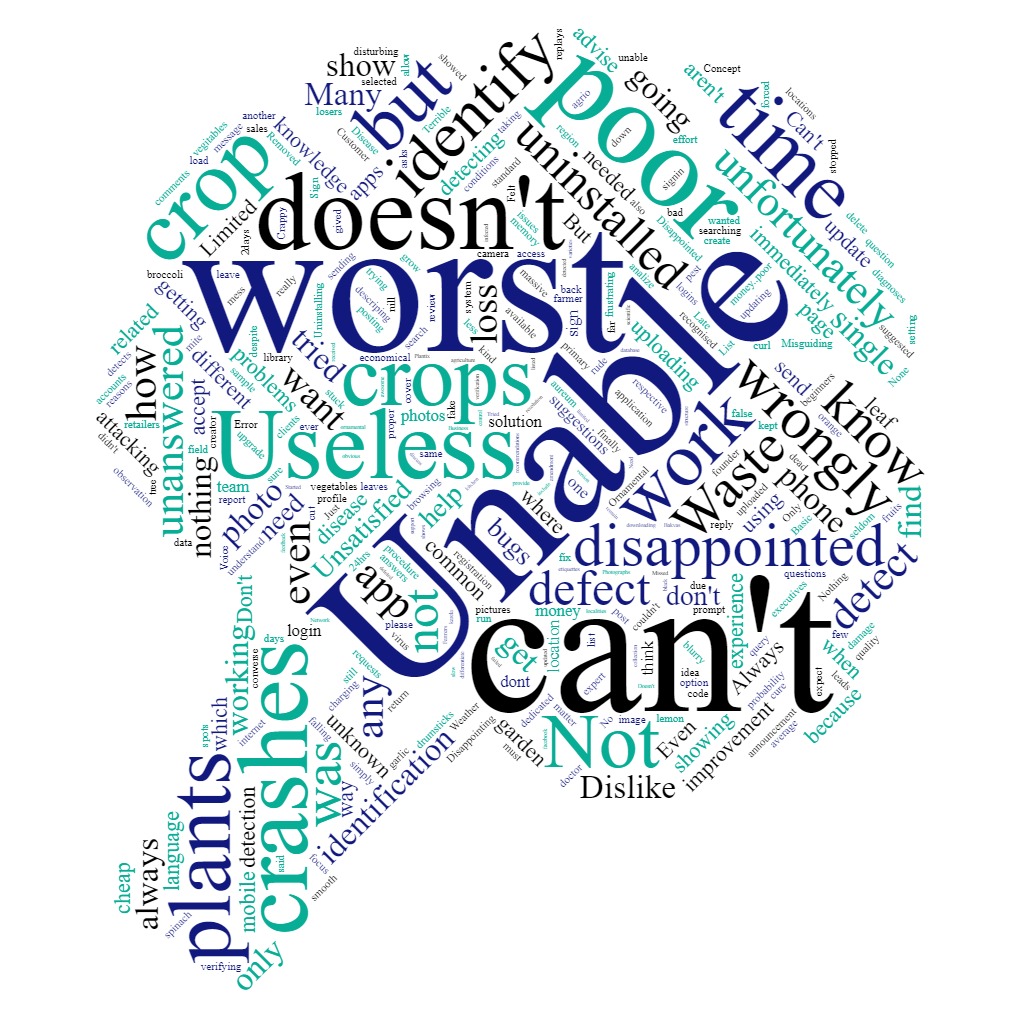}
    \caption{Negative comments}
    \label{bad}
  \end{subfigure}
  \caption{Word-cloud of positive and negative comments.}
  \label{word cloud}
\end{figure}

In this research, to understand the overall user experience of those apps, two word-clouds have been created, as shown in Fig.~\ref{word cloud}. Word-clouds are graphical representations of word frequency. The larger the word in the graphic, the more frequently that word appear in the text~\cite{prowritingaid}.

User comments from the app listing from the respective app stores along with the app metadata were collected for this study. Based on the users' review ratings of the app, the consumer comment sections were split into two kinds - the comment is positive if its rating is 4 stars or above, otherwise it is considered negative. The study of user feedback from both app stores on plant disease detection apps reveals that most apps have little or no strong feedback or favorable ratings from smartphone users because the count of apps installed was generally poor for most of the relevant applications. Moreover, Apple App store does not show the number of downloads of an app. 

However, 35.30\% (6/17) of the apps contain informative comments and have 10K+ to 10M+ installations. A word-cloud is generated in Fig.~\ref{good} using the positive comments collected from the selected apps. One of the most optimistic collections of user comments has been noticed for \app{Plantix-your crop doctor}.  This app has already received 60K+ comments, with 65\% being 5-star reviews. Among the 35K+ highest star rating comments, people mentioned the good points of this app. Users have used several words to express their satisfaction, such as ``helpful", ``wonderful", ``amazing", ``fantastic", ``awesome", ``very good", ``great", ``best", ``excellent", ``useful", ``easy to use", ``accurate", ``nice" and many more. Many users gave ``thanks" and praised the expert team of this app community. 
According to some users, the finest plant disease diagnosis app is \app{Plantix - your crop doctor}. The app has several wonderful features such as detection ability for crop diseases, a fertilizer calculator, and pest and disease-related information with high-resolution pictures. The app provides a community support where plant lovers can share plant-related problems and get suggestions from other users and experts. One user has commented  ``\textit{Hello team plantix You people are doing wonderful job... I personally loved this app this app gives complete guidance for farming a variety of crops. Gives details not only about disease but also about its treatment also provide guidance about fertilization of crop. It would be great if more people install and give it a shot you wont regret of installing it . Best luck team plantix and thanks a lot.}". This comment is supported by many users. Another app, \app{Agrio}, has received the second-highest number of positive reviews from users.

A word-cloud is generated in Fig.~\ref{bad} using the negative comments collected from the selected apps. Some users provided their damaged or diseased leaves images to the app, but could not detect the disease, or sometimes even the plants, prompting some comments to include the word ``unable". The word ``stuck” is used to indicate that an app had been trying to load a certain step or was being so sluggish that it seemed to have halted its task. Some apps have a lengthy lag, most often unable to load an image from their servers, so they can take a considerable time to load affected crop images. People use the words ``unfortunately" and ``worst" to express their dissatisfaction in these situations. Some apps' databases are limited and require more data. Sometimes apps show an unexpected error message. Other problems include app ``crashes” where the site does not function correctly. That is why users use the word ``useless” many times.

Users' dissatisfaction related to the \app{Plant Disease Detection} and \app{PlantifyDr} apps was focused on several points, including limited crop and plant availability, an inability to detect plants, failure to login, the system displaying unknown errors, incorrect weather reports, and failure to capture any pictures with a high-resolution camera. A user posted a comment that was supported by 1k+ other users. He wrote: ``\textit{Useless app for me. The app asks you to pick the top 4 crops you are interested in and it is really frustrating that you are forced to select from a VERY limited collection of crops that don’t include the ones I want, like kale, spinach, garlic, broccoli, beets.}"

The \app{Leaf Doctor} app can not always determine the correct threshold of the given leaf. It indicates inaccurate thresholds in most situations. \app{Riceye} provides garbage threshold every time. When a picture is imported into the \app{PlantDoctor} app, it always responds that the image is unclear, despite the fact even if the image is in HD quality. The \app{Pestoz} app has poor service and needs major upgrades. It has a limited database of diseased plant images. Sometimes it shows an unexpected error message. It ``crashes" and the site does not function correctly. \app{Plantix - your crop doctor} covers 45+ crops and 600+ diseases related to them, does not support all ornamental plants and also has a community to share queries with experts. The \app{Agrio} app is sometimes stuck on the home interface. Consumers used the ``worst" word several times to describe these incidents. Users put their damaged crops to the ``Agrio community" but no help or suggestion was found from the team. The \app{Cropalyser} app takes a considerable time to load crop images and is most often unable to load an image from its servers. \app{PDDApp: Plant Disease Detection} shows a ``Check Connection" message every time through email, even with a strong internet connection, it can not run a single job. \app{Plant Disease Detector} is a flawed app that can not detect plants or leaves as it shows two random thresholds of any thing.

In summary, most of the currently available apps suffer from serious system design flaws and development issues and need improvement.

\section{Discussions}\label{sec:discussions}
\subsection{Principal findings}\label{sec:PriFind} 

We found many apps in the app stores when we searched by the keywords ``plant disease", ``leaf disease detection", etc., but only a few of them were well designed and developed to successfully detect plant diseases. Unfortunately, those few apps of key interest also lacked the seven basic functionalities related to plant disease detection. 
Both plant identification and disease detection are present in 88.23\% (15/17) of the apps. While selecting a plant from a pre-specified category, sometime apps cannot identify it correctly. Plant disease detection is the central feature of all these apps, yet \app{Leaf Doctor} and \app{PlantDoctor}, do not provide this feature. \app{PlantDoctor} gives the option of taking pictures through the device camera and every time it shows the message ``The image is not clear to be identified” even for a high quality image. Another app, \app{Plant Disease Detector}, gives a verdict for both diseased and healthy plants by saying only ``Diseased”. Severity assessment helps the farmers to decide whether their plant needs treatment and to what extent. This feature is present in only 23.52\% (4/17) of the apps, where two of them (\app{Plant Disease Detector} and \app{Riceye}) display a random threshold value. 

We have noticed that some apps' designs are troublesome. First, 35\% (6/17) of the apps need an uninterrupted internet connection to provide the complete service, which is difficult to manage for village farmers or users living in remote areas where the internet service is often poor. 
Second, a handful of apps cannot upload pictures from the phone photo gallery, causing problems for users with low quality device cameras or who may not have the proper illumination to take clear pictures while using the app. Third, most of the apps we reviewed were not multilingual, and thus, users of these apps must have a basic knowledge of English.

We have found several beneficial features in some of the reviewed apps. \app{Plantix - your crop doctor} is the best for plant disease identification. This app detects pests and diseases on crops and recommends the necessary treatment, gives disease alerts, and contains a fertilizer calculator for crops based on the farmer's plot measurements. It also gives a weather forecast to let gardeners know the best time to weed, spray and harvest along with cultivation tips. Users can interact with researchers, farmers, and plant professionals in an online forum to share plant health issues. They have more than 30 million images in their data set, which contains more than 400 crop diseases. This app applies ML algorithms and AI to photos to diagnose crops. It also allows for real-time disease and pest monitoring. The user's images are submitted to the servers and are immediately examined to identify the disorder. The user is given essential information about the symptoms, triggers, agrochemicals, and biological remedies. The app includes a meteorological information system customized to farmers' locations, and a community option that allows them to communicate with other plant protection groups and thus, increases productivity~\cite{CASESTUDY}. 

Another app, \app{Leaf doctor}, is slightly advanced in that users can select healthy tissues using up to eight colors in the photograph and then be provided with the disease severity estimation. Also, users can fix the threshold by using a slider provided in the app. A significant flaw is that this app gives the severity estimation for any type of object which is not necessarily related to plants or leaves. A unique feature provided by the app \app{AgroAI - Plant Diseases Diagnosis (Early Access)} is the soil fertility test, where landowners can input the pH value of soil and see the proportion of different soil elements such as calcium, potassium, clay, silt, etc. but this feature needs more development. Another app with only the plant disease identification feature is \app{Leafy}, which works offline and identifies the correct plant and its disease, though it has a limited diseases list (only 30). 

Providing treatment or prevention alongside disease detection is a helpful feature provided by 64.71\% (11/17) of the apps. Also, many apps shared disease information with accompanying disease images, which is another useful feature for information gathering related to plant diseases. Community help is provided by two apps \app{Plantix - your crop doctor} and \app{Agrio}. In the \app{Plantix - your crop doctor} community, all gardeners, botanists and agricultural experts can communicate and exchange suggestions - this is highly beneficial. However, in the \app{Agrio} community, plant disease detection is manually handled by their own expert team commenting in posts and suggesting treatments. Moreover, any other people can also comment in the community forum. This is highly beneficial for increasing farmers' knowledge, improving their attitudes toward plant management, and increasing their help-seeking mentality. However, the number of such apps is minimal. Most of the apps lack basic functionality and do not benefit the users. 

Automatic visualization of disease in an infected leaf helps to identify the diseased portion of a plant and this helpful functionality is missing from all the apps. The existing apps need their basic features improvement where basic features will be present andd to they will work more accurately.

\subsection{Design considerations}

There are many aspects of existing plant disease detection apps that require significant technical improvements. A simple yet attractive user interface with high-resolution graphics will attract users and increase their app exploring time. Multilingual apps will encourage users from different countries to use the app and also give a better understanding of the apps' functionalities. Apps that provide tutorials at the beginning will help users operate the app easily and correctly. Another simple but effective feature of an app that will keep users up to date on the status of their plants is regular notifications. Any efficient app will have low battery power consumption and low memory usage. Many of our reviewed apps frequently crashed while in use. This issue should be considered while developing apps, as it has a large negative impact on users. Also, the user interface components should work smoothly and be consistent.

From the functionality perspective, apps should perform all the operations accurately. Apps that cannot identify plants can keep major categories of plants such as crops, vegetables, ornamental plants, medicinal plants, etc., covering many different plant types so that users can easily find their plants. The most important feature is plant disease detection, which is the apps' focus. Apps should detect the diseases accurately and, in the case of similar diseases, they should be able to give the likelihood of each disease. Disease severity detection is another basic criterion to understand the threat level of the disease that app should provide. To perform the disease severity estimation properly, visualizing the infected area of the diseased portion is a prerequisite. Therefore, visualization has become a crucial task that should be performed by the apps using existing image processing and ML technologies. Additionally, more information about the disease and its control measures benefits users, which encourages them to take further disease management steps from the apps.

\subsection{Limitations of this study}
Several apps were unfortunately discarded due to regional restrictions, specific language restrictions, and the need to have a license key to use the app. Furthermore, since our app search and evaluation, new apps may have been added, and some of the apps might have been updated with enhanced functionality, or been removed from the app stores. In this paper, we did not evaluate the accuracy of the disease detection performed by those reviewed apps but rather focused on the software quality aspects, app features, and to what extent artificial intelligence is being adopted to provide advanced disease detection functionality.

\section{Conclusion and future work}\label{sec:conclusion}
This paper features a survey on plant disease detection apps from three popular app stores. Our search results identified a total of 606 apps, from which we selected 17 based on a defined set of inclusion and exclusion criteria. We further evaluated those 17 apps using our devised app rating scale. Our app rating scale for plant disease detection evaluated the AI-based functionality and software quality of apps. 
According to our findings, most of the apps fell significantly below standard. They tended to lack many basic and important functionalities to detect plant disease, and many of them give users an inappropriate or incorrect verdict or solution. Although a few apps had some of the expected features, none of those under review met all the required functionalities. We also observed that there were very few evidence-based apps. As there have been numerous studies about plant disease detection and identification, disease severity estimation, localization and visualization, such technological advancements must be included in modern plant disease detection apps.
We hope that the developers mitigate the shortcomings found in existing apps so that the upcoming or updated apps will contain suggested new functionalities and perform those features correctly. Producers and botanists will soon be able to use these improved plant disease identification apps in the field. Users will be inspired to use these apps for detection of any type of disease or plant pathogen, while others will be motivated to use these apps for further plant management assistance.

In the future, we plan to address the identified limitations and study numerous apps from different app stores. Also, our rating scale will be extended by incorporating the accuracy metrics related to plant disease detection. We will conduct a formal sentiment analysis on user reviews in app stores. We will also improve the reviewer panel so that different aspects of the apps, such as software quality and plant-specific functionalities, get appropriate attention and improved evaluation.


\bibliographystyle{unsrt}  
\bibliography{references}

\begin{thebibliography}{10}

\bibitem{agrios2005plant}
George~N Agrios.
\newblock {\em Plant pathology}.
\newblock Academic Press, fifth edition edition, 2005.

\bibitem{leafDoctor}
Malusi Sibiya and Mbuyu Sumbwanyambe.
\newblock An algorithm for severity estimation of plant leaf diseases by the
  use of colour threshold image segmentation and fuzzy logic inference: a
  proposed algorithm to update a ``leaf doctor" application.
\newblock {\em AgriEngineering}, 1(2):205--219, 2019.

\bibitem{plantdisease}
Malcolm~C. Shurtleff, Rita~M. Pelczar, Michael~J. Pelczar, and Arthur Kelman.
\newblock Plant disease.
\newblock {\em Encyclopedia Britannica}, 2021.

\bibitem{phadikar2008rice}
Santanu Phadikar and Jaya Sil.
\newblock Rice disease identification using pattern recognition techniques.
\newblock In {\em 2008 11th International Conference on Computer and
  Information Technology}, pages 420--423. IEEE, 2008.

\bibitem{islam2017detection}
Monzurul Islam, Anh Dinh, Khan Wahid, and Pankaj Bhowmik.
\newblock Detection of potato diseases using image segmentation and multiclass
  support vector machine.
\newblock In {\em IEEE 30th canadian conference on electrical and computer
  engineering (CCECE)}, pages 1--4. IEEE, 2017.

\bibitem{huang2014new}
Wenjiang Huang, Qingsong Guan, Juhua Luo, Jingcheng Zhang, Jinling Zhao, Dong
  Liang, Linsheng Huang, and Dongyan Zhang.
\newblock New optimized spectral indices for identifying and monitoring winter
  wheat diseases.
\newblock {\em IEEE Journal of Selected Topics in Applied Earth Observations
  and Remote Sensing}, 7(6):2516--2524, 2014.

\bibitem{article}
Sharma Ram, Kumarse Nazari, Amir Amanov, Zafarjon Ziyaev, and Anwar Jalilov.
\newblock Reduction of winter wheat yield losses caused by stripe rust through
  fungicide management.
\newblock {\em Journal of Phytopathology}, 164, 2016.

\bibitem{meunkaewjinda2008grape}
A~Meunkaewjinda, P~Kumsawat, K~Attakitmongcol, and A~Srikaew.
\newblock Grape leaf disease detection from color imagery using hybrid
  intelligent system.
\newblock In {\em 5th international conference on electrical
  engineering/electronics, computer, telecommunications and information
  technology}, volume~1, pages 513--516. IEEE, 2008.

\bibitem{sannakki2013diagnosis}
Sanjeev~S Sannakki, Vijay~S Rajpurohit, VB~Nargund, and Pallavi Kulkarni.
\newblock Diagnosis and classification of grape leaf diseases using neural
  networks.
\newblock In {\em 2013 Fourth International Conference on Computing,
  Communications and Networking Technologies (ICCCNT)}, pages 1--5. IEEE, 2013.

\bibitem{6707647}
Monika Jhuria, Ashwani Kumar, and Rushikesh Borse.
\newblock Image processing for smart farming: Detection of disease and fruit
  grading.
\newblock In {\em 2013 IEEE Second International Conference on Image
  Information Processing (ICIIP-2013)}, pages 521--526, 2013.

\bibitem{Xie}
Xiaoyue Xie, Yuan Ma, Liu Bin, Jinrong He, Shuqin Li, and Hongyan Wang.
\newblock A deep-learning-based real-time detector for grape leaf diseases
  using improved convolutional neural networks.
\newblock {\em Frontiers in Plant Science}, 11, 2020.

\bibitem{KARLEKAR2020105342}
Aditya Karlekar and Ayan Seal.
\newblock Soynet: Soybean leaf diseases classification.
\newblock {\em Computers and Electronics in Agriculture}, 172:105342, 2020.

\bibitem{al2011fast}
Heba Al-Hiary, Sulieman Bani-Ahmad, M~Reyalat, Malik Braik, and Zainab
  Alrahamneh.
\newblock Fast and accurate detection and classification of plant diseases.
\newblock {\em International Journal of Computer Applications}, 17(1):31--38,
  2011.

\bibitem{khirade2015plant}
Sachin~D Khirade and AB~Patil.
\newblock Plant disease detection using image processing.
\newblock In {\em 2015 International conference on computing communication
  control and automation}, pages 768--771. IEEE, 2015.

\bibitem{cassava}
Amanda Ramcharan, Peter McCloskey, Kelsee Baranowski, Neema Mbilinyi, Latifa
  Mrisho, Mathias Ndalahwa, James Legg, and David~P Hughes.
\newblock A mobile-based deep learning model for cassava disease diagnosis.
\newblock {\em Frontiers in plant science}, 10:272, 2019.

\bibitem{Dutot}
M.~Dutot, L.~Nelson, and Rebecca Tyson.
\newblock Predicting the spread of postharvest disease in stored fruit, with
  application to apples.
\newblock {\em Postharvest Biology and Technology}, 85:45--56, 11 2013.

\bibitem{francis2016identification}
Jobin Francis, BK~Anoop, et~al.
\newblock Identification of leaf diseases in pepper plants using soft computing
  techniques.
\newblock In {\em 2016 conference on emerging devices and smart systems
  (ICEDSS)}, pages 168--173. IEEE, 2016.

\bibitem{petrellis2017mobile}
Nikos Petrellis.
\newblock Mobile application for plant disease classification based on symptom
  signatures.
\newblock In {\em 21st Pan-Hellenic Conference on Informatics}, pages 1--6,
  2017.

\bibitem{riley2002plant}
Melissa~B Riley, Margaret~R Williamson, and Otis Maloy.
\newblock Plant disease diagnosis.
\newblock {\em The Plant Health Instructor}, 10, 2002.

\bibitem{BEDI202190}
Punam Bedi and Pushkar Gole.
\newblock Plant disease detection using hybrid model based on convolutional
  autoencoder and convolutional neural network.
\newblock {\em Artificial Intelligence in Agriculture}, 5:90--101, 2021.

\bibitem{agriengineering3030032}
Ahmed~Abdelmoamen Ahmed and Gopireddy~Harshavardhan Reddy.
\newblock A mobile-based system for detecting plant leaf diseases using deep
  learning.
\newblock {\em AgriEngineering}, 3(3):478--493, 2021.

\bibitem{ict}
Shitala Prasad, Sateesh~K Peddoju, and Debashis Ghosh.
\newblock Agromobile: a cloud-based framework for agriculturists on mobile
  platform.
\newblock {\em International Journal of advanced science and technology},
  59:41--52, 2013.

\bibitem{singh2018deep}
Asheesh~Kumar Singh, Baskar Ganapathysubramanian, Soumik Sarkar, and Arti
  Singh.
\newblock Deep learning for plant stress phenotyping: trends and future
  perspectives.
\newblock {\em Trends in plant science}, 23(10):883--898, 2018.

\bibitem{9399342}
Lili Li, Shujuan Zhang, and Bin Wang.
\newblock Plant disease detection and classification by deep learning—a
  review.
\newblock {\em IEEE Access}, 9:56683--56698, 2021.

\bibitem{cottonfarming}
Cotton Farming.
\newblock Plant disease detection using smartphone goes high tech.
\newblock
  \url{https://www.cottonfarming.com/production-2/plant-disease-detection-using-smartphone-goes-high-tech/},
  2019.
\newblock (accessed 10 June 2021).

\bibitem{petrellis2019plant}
Nikos Petrellis.
\newblock Plant disease diagnosis for smart phone applications with extensible
  set of diseases.
\newblock {\em Applied Sciences}, 9(9):1952, 2019.

\bibitem{ashok2020tomato}
Surampalli Ashok, Gemini Kishore, Velpula Rajesh, S~Suchitra, SG~Gino Sophia,
  and B~Pavithra.
\newblock Tomato leaf disease detection using deep learning techniques.
\newblock In {\em 2020 5th International Conference on Communication and
  Electronics Systems (ICCES)}, pages 979--983. IEEE, 2020.

\bibitem{weizheng2008grading}
Shen Weizheng, Wu~Yachun, Chen Zhanliang, and Wei Hongda.
\newblock Grading method of leaf spot disease based on image processing.
\newblock In {\em 2008 international conference on computer science and
  software engineering}, volume~6, pages 491--494. IEEE, 2008.

\bibitem{5697452}
Dheeb Al~Bashish, Malik Braik, and Sulieman Bani-Ahmad.
\newblock A framework for detection and classification of plant leaf and stem
  diseases.
\newblock In {\em 2010 International Conference on Signal and Image
  Processing}, pages 113--118, 2010.

\bibitem{arsenovic2019solving}
Marko Arsenovic, Mirjana Karanovic, Srdjan Sladojevic, Andras Anderla, and
  Darko Stefanovic.
\newblock Solving current limitations of deep learning based approaches for
  plant disease detection.
\newblock {\em Symmetry}, 11(7):939, 2019.

\bibitem{agriculture12010009}
Houda Orchi, Mohamed Sadik, and Mohammed Khaldoun.
\newblock On using artificial intelligence and the internet of things for crop
  disease detection: A contemporary survey.
\newblock {\em Agriculture}, 12(1), 2022.

\bibitem{PICON2019280}
Artzai Picon, Aitor Alvarez-Gila, Maximiliam Seitz, Amaia Ortiz-Barredo, Jone
  Echazarra, and Alexander Johannes.
\newblock Deep convolutional neural networks for mobile capture device-based
  crop disease classification in the wild.
\newblock {\em Computers and Electronics in Agriculture}, 161:280--290, 2019.

\bibitem{YUAN202248}
Yuan Yuan, Lei Chen, Huarui Wu, and Lin Li.
\newblock Advanced agricultural disease image recognition technologies: A
  review.
\newblock {\em Information Processing in Agriculture}, 9(1):48--59, 2022.

\bibitem{che2022mobile}
Nik~Norasma Che’Ya, Nur~Adibah Mohidem, Nor~Athirah Roslin, Mohammadmehdi
  Saberioon, Mohammad~Zakri Tarmidi, Jasmin Arif~Shah, Wan~Fazilah
  Fazlil~Ilahi, and Norsida Man.
\newblock Mobile computing for pest and disease management using spectral
  signature analysis: A review.
\newblock {\em Agronomy}, 12(4):967, 2022.

\bibitem{sibanda2021systematic}
Blessing~K Sibanda, Gloria~E Iyawa, and Attlee~M Gamundani.
\newblock Systematic review of plant pest and disease identification strategies
  and techniques in mobile apps.
\newblock In {\em World Conference on Information Systems and Technologies},
  pages 491--502. Springer, 2021.

\bibitem{Agrisurvey}
Hetal Patel and Dharmendra Patel.
\newblock Survey of android apps for agriculture sector.
\newblock {\em International Journal of Information Sciences and Techniques},
  6(1-2):61--67, 2016.

\bibitem{multimediaCrop}
{Bhavsar, Shreyaben. B.}
\newblock {\em Multilingual multimedia based crop disease management system
  using mobile technology}.
\newblock PhD thesis, Sardar Patel University, Gujarat, India, 2019.

\bibitem{rivera2016mobile}
Jordan Rivera, Amy McPherson, Jill Hamilton, Catherine Birken, Michael Coons,
  Sindoora Iyer, Arnav Agarwal, Chitra Lalloo, and Jennifer Stinson.
\newblock Mobile apps for weight management: a scoping review.
\newblock {\em JMIR mHealth and uHealth}, 4(3):e5115, 2016.

\bibitem{Kabir2021}
Muhammad~Ashad Kabir, Sheikh~Sowmen Rahman, Mohammad~Mainul Islam, Sayed Ahmed,
  and Craig Laird.
\newblock Mobile apps for foot measurement in pedorthic practice: Scoping
  review.
\newblock {\em JMIR Mhealth Uhealth}, 9(3), Mar 2021.

\bibitem{tricco2018prisma}
Andrea~C Tricco, Erin Lillie, Wasifa Zarin, Kelly~K O'Brien, Heather Colquhoun,
  Danielle Levac, David Moher, Micah~DJ Peters, Tanya Horsley, Laura Weeks,
  et~al.
\newblock Prisma extension for scoping reviews (prisma-scr): checklist and
  explanation.
\newblock {\em Annals of internal medicine}, 169(7):467--473, 2018.

\bibitem{liberati2009prisma}
Alessandro Liberati, Douglas~G Altman, Jennifer Tetzlaff, Cynthia Mulrow,
  Peter~C G{\o}tzsche, John~PA Ioannidis, Mike Clarke, Philip~J Devereaux, Jos
  Kleijnen, and David Moher.
\newblock The prisma statement for reporting systematic reviews and
  meta-analyses of studies that evaluate health care interventions: explanation
  and elaboration.
\newblock {\em Journal of clinical epidemiology}, 62(10):e1--e34, 2009.

\bibitem{stoyanov2015mobile}
Stoyan~R Stoyanov, Leanne Hides, David~J Kavanagh, Oksana Zelenko, Dian
  Tjondronegoro, and Madhavan Mani.
\newblock Mobile app rating scale: a new tool for assessing the quality of
  health mobile apps.
\newblock {\em JMIR mHealth and uHealth}, 3(1):e27, 2015.

\bibitem{pritha2022smartphone}
Sadia~Tasnuva Pritha, Rahnuma Tasnim, Ashad Kabir, Sumaiya Amin, and Anik Das.
\newblock Smartphone apps for child sexual abuse education: Gaps and design
  considerations.
\newblock {\em International Journal of Mobile Learning and Organisation},
  2022.

\bibitem{sabiha2022}
Sabiha Samad, Fahmida Ahmed, Samsun Naher, Muhammad~Ashad Kabir, Anik Das,
  Sumaiya Amin, and Sheikh Mohammed~Shariful Islam.
\newblock Smartphone apps for tracking food consumption and recommendations:
  Evaluating artificial intelligence-based functionalities, features and
  quality of current apps.
\newblock {\em Intelligent Systems with Applications}, 15:200103, 2022.

\bibitem{umars}
Stoyan~R Stoyanov, Leanne Hides, David~J Kavanagh, and Hollie Wilson.
\newblock Development and validation of the user version of the mobile
  application rating scale (umars).
\newblock {\em JMIR mHealth and uHealth}, 4(2):e72, 2016.

\bibitem{finmars}
Johannes Huebner, Carlo Schmid, Mehdi Bouguerra, and Alexander Ilic.
\newblock Finmars: A mobile app rating scale for finance apps.
\newblock In {\em 9th International Conference on Information Communication and
  Management}, pages 6--11, 2019.

\bibitem{wu2007empirical}
Chien-Ho Wu.
\newblock An empirical study on the transformation of likert-scale data to
  numerical scores.
\newblock {\em Applied Mathematical Sciences}, 1(58):2851--2862, 2007.

\bibitem{silvennoinen2014experiencing}
Experiencing visual usability and aesthetics in two mobile application
  contexts.
\newblock {\em Journal of usability studies}, 10(1), 2014.

\bibitem{von2018we}
Christiane~G von Wangenheim, Jo{\~a}o V~Araujo Porto, Jean~CR Hauck, and
  Adriano~F Borgatto.
\newblock Do we agree on user interface aesthetics of android apps?
\newblock {\em arXiv preprint arXiv:1812.09049}, 2018.

\bibitem{isO}
ISO.
\newblock Iso/iec 25010:2011 systems and software engineering — systems and
  software quality requirements and evaluation (square) — system and software
  quality models.
\newblock \url{https://www.iso.org/standard/35733.html}, 2011.
\newblock (accessed 10 June 2021).

\bibitem{zen2016assessing}
Mathieu Zen and Jean Vanderdonckt.
\newblock Assessing user interface aesthetics based on the inter-subjectivity
  of judgment.
\newblock In {\em 30th International BCS Human Computer Interaction Conference
  30}, pages 1--12, 2016.

\bibitem{tractinsky2000beautiful}
Noam Tractinsky, Adi~S Katz, and Dror Ikar.
\newblock What is beautiful is usable.
\newblock {\em Interacting with computers}, 13(2):127--145, 2000.

\bibitem{sensortower}
Stephanie Chan.
\newblock 5-year market forecast: App spending will climb to \$270 billion by
  2025.
\newblock
  \url{https://sensortower.com/blog/sensor-tower-app-market-forecast-2025},
  2021.
\newblock (accessed 15 July 2021).

\bibitem{wei2015usability}
Qunyi Wei, Zhaoxin Chang, and Qin Cheng.
\newblock Usability study of the mobile library app: an example from chongqing
  university.
\newblock {\em Library Hi Tech}, 2015.

\bibitem{payne2015behavioral}
Hannah~E Payne, Cameron Lister, Joshua~H West, and Jay~M Bernhardt.
\newblock Behavioral functionality of mobile apps in health interventions: a
  systematic review of the literature.
\newblock {\em JMIR mHealth and uHealth}, 3(1):e3335, 2015.

\bibitem{kallio2005usability}
Titti Kallio, Anne Kaikkonen, et~al.
\newblock Usability testing of mobile applications: A comparison between
  laboratory and field testing.
\newblock {\em Journal of Usability studies}, 1(4-16):23--28, 2005.

\bibitem{7961668}
Maleknaz Nayebi, Mahshid Marbouti, Rache Quapp, Frank Maurer, and Guenther
  Ruhe.
\newblock Crowdsourced exploration of mobile app features: A case study of the
  fort mcmurray wildfire.
\newblock In {\em IEEE/ACM 39th International Conference on Software
  Engineering: Software Engineering in Society Track (ICSE-SEIS)}, pages
  57--66, 2017.

\bibitem{miller2009plant}
Sally~A Miller, Fen~D Beed, and Carrie~Lapaire Harmon.
\newblock Plant disease diagnostic capabilities and networks.
\newblock {\em Annual review of phytopathology}, 47:15--38, 2009.

\bibitem{mahlein2016plant}
Anne-Katrin Mahlein.
\newblock Plant disease detection by imaging sensors--parallels and specific
  demands for precision agriculture and plant phenotyping.
\newblock {\em Plant disease}, 100(2):241--251, 2016.

\bibitem{waldchen2018automated}
Jana W{\"a}ldchen, Michael Rzanny, Marco Seeland, and Patrick M{\"a}der.
\newblock Automated plant species identification—trends and future
  directions.
\newblock {\em PLoS computational biology}, 14(4):e1005993, 2018.

\bibitem{wang2017review}
Zhaobin Wang, Huale Li, Ying Zhu, and TianFang Xu.
\newblock Review of plant identification based on image processing.
\newblock {\em Archives of Computational Methods in Engineering},
  24(3):637--654, 2017.

\bibitem{nijalingappa2015plant}
Pradeep Nijalingappa and VJ~Madhumathi.
\newblock Plant identification system using its leaf features.
\newblock In {\em 2015 International Conference on Applied and Theoretical
  Computing and Communication Technology (iCATccT)}, pages 338--343. IEEE,
  2015.

\bibitem{agarwal2018plant}
Sonali Agarwal, Anand~Singh Jalal, Mohd Khan, et~al.
\newblock Plant identification using leaf image analysis.
\newblock In {\em 3rd International Conference on Internet of Things and
  Connected Technologies (ICIoTCT)}, pages 26--27, 2018.

\bibitem{vanrecognition}
Ngo Le~Huy~Hien and Nguyen V.H.
\newblock Recognition of plant species using deep convolutional feature
  extraction.
\newblock {\em International Journal on Emerging Technologies}, 11:904--910, 06
  2020.

\bibitem{bodhwani2019deep}
Vinit Bodhwani, DP~Acharjya, and Umesh Bodhwani.
\newblock Deep residual networks for plant identification.
\newblock {\em Procedia Computer Science}, 152:186--194, 2019.

\bibitem{FERENTINOS2018311}
Konstantinos~P. Ferentinos.
\newblock Deep learning models for plant disease detection and diagnosis.
\newblock {\em Computers and Electronics in Agriculture}, 145:311--318, 2018.

\bibitem{Plantvillage}
David Hughes and Marcel Salath{\'e}.
\newblock An open access repository of images on plant health to enable the
  development of mobile disease diagnostics.
\newblock {\em arXiv preprint arXiv:1511.08060}, 2015.

\bibitem{geetharamani2019identification}
G~Geetharamani and Arun Pandian.
\newblock Identification of plant leaf diseases using a nine-layer deep
  convolutional neural network.
\newblock {\em Computers \& Electrical Engineering}, 76:323--338, 2019.

\bibitem{ABADE2021106125}
André Abade, Paulo~Afonso Ferreira, and Flavio {de Barros Vidal}.
\newblock Plant diseases recognition on images using convolutional neural
  networks: A systematic review.
\newblock {\em Computers and Electronics in Agriculture}, 185:106125, 2021.

\bibitem{BISCHOFF2021105922}
Vinicius Bischoff, Kleinner Farias, Juliano~Paulo Menzen, and Gustavo Pessin.
\newblock Technological support for detection and prediction of plant diseases:
  A systematic mapping study.
\newblock {\em Computers and Electronics in Agriculture}, 181:105922, 2021.

\bibitem{LEE2020105220}
Sue~Han Lee, Hervé Goëau, Pierre Bonnet, and Alexis Joly.
\newblock New perspectives on plant disease characterization based on deep
  learning.
\newblock {\em Computers and Electronics in Agriculture}, 170:105220, 2020.

\bibitem{WSPANIALY2020105701}
Patrick Wspanialy and Medhat Moussa.
\newblock A detection and severity estimation system for generic diseases of
  tomato greenhouse plants.
\newblock {\em Computers and Electronics in Agriculture}, 178:105701, 2020.

\bibitem{arivazhagan2013detection}
Sai Arivazhagan, R~Newlin Shebiah, S~Ananthi, and S~Vishnu Varthini.
\newblock Detection of unhealthy region of plant leaves and classification of
  plant leaf diseases using texture features.
\newblock {\em Agricultural Engineering International: CIGR Journal},
  15(1):211--217, 2013.

\bibitem{LIANG2019518}
Qiaokang Liang, Shao Xiang, Yucheng Hu, Gianmarc Coppola, Dan Zhang, and Wei
  Sun.
\newblock Pd2se-net: Computer-assisted plant disease diagnosis and severity
  estimation network.
\newblock {\em Computers and Electronics in Agriculture}, 157:518--529, 2019.

\bibitem{betzing2020impact}
Jan~Hendrik Betzing, Matthias Tietz, Jan vom Brocke, and J{\"o}rg Becker.
\newblock The impact of transparency on mobile privacy decision making.
\newblock {\em Electronic Markets}, 30(3):607--625, 2020.

\bibitem{HotMobile}
Alastair Beresford, Andrew Rice, Nicholas Skehin, and Ripduman Sohan.
\newblock Mockdroid: Trading privacy for application functionality on
  smartphones.
\newblock {\em HotMobile 2011: The 12th Workshop on Mobile Computing Systems
  and Applications}, 03 2011.

\bibitem{christmann2006robust}
Andreas Christmann and Stefan Van~Aelst.
\newblock Robust estimation of cronbach's alpha.
\newblock {\em Journal of Multivariate Analysis}, 97(7):1660--1674, 2006.

\bibitem{cronbach1951coefficient}
Lee~J Cronbach.
\newblock Coefficient alpha and the internal structure of tests.
\newblock {\em psychometrika}, 16(3):297--334, 1951.

\bibitem{gliem2003calculating}
Joseph Gliem and Rosemary Gliem.
\newblock Calculating, interpreting, and reporting cronbach’s alpha
  reliability coefficient for likert-type scales.
\newblock {\em 2003 Midwest Research to Practice Conference in Adult,
  Continuing, and Community Education}, 2003.

\bibitem{lange2011inter}
RT~Lange.
\newblock Inter-rater reliability.
\newblock {\em Encyclopedia of Clinical Neuropsychology. New York, NY: Springer
  New York}, page 1348, 2011.

\bibitem{sawa2007interrater}
Jungo Sawa and Toshihiko Morikawa.
\newblock Interrater reliability for multiple raters in clinical trials of
  ordinal scale.
\newblock {\em Drug information journal: DIJ/Drug Information Association},
  41(5):595--605, 2007.

\bibitem{koo2016guideline}
Terry~K Koo and Mae~Y Li.
\newblock A guideline of selecting and reporting intraclass correlation
  coefficients for reliability research.
\newblock {\em Journal of chiropractic medicine}, 15(2):155--163, 2016.

\bibitem{pethybridge2015leaf}
Sarah~J Pethybridge and Scot~C Nelson.
\newblock Leaf doctor: A new portable application for quantifying plant disease
  severity.
\newblock {\em Plant disease}, 99(10):1310--1316, 2015.

\bibitem{6636712}
Dennis Pagano and Walid Maalej.
\newblock User feedback in the appstore: An empirical study.
\newblock In {\em 2013 21st IEEE International Requirements Engineering
  Conference (RE)}, pages 125--134, 2013.

\bibitem{prowritingaid}
Lisa Lepki.
\newblock How to use word clouds for business, fiction and copywriting.
\newblock
  \url{https://prowritingaid.com/art/425/What-the-Heck-is-a-Word-Cloud-and-Why-Would-I-Use-One.aspx},
  2020.
\newblock (accessed 5 July 2021).

\bibitem{CASESTUDY}
CGIAR.
\newblock Plant disease diagnosis using artificial intelligence: A case study
  on plantix.
\newblock
  \url{https://bigdata.cgiar.org/digital-intervention/plant-disease-diagnosis-using-artificial-intelligence-a-case-study-on-plantix/},
  2021.
\newblock (accessed 20 October 2021).

\end{thebibliography}

\end{document}